\documentclass[12pt,preprint]{aastex}

\slugcomment{Astrophysical Journal}
\shorttitle{Surface photon emissivity of bare strange stars}
\shortauthors{K.S.~Cheng and T.~Harko}
\begin{document}
\title{Surface photon emissivity of bare strange stars}
\author{K.S.~Cheng $^1$ and T.~Harko$^2$}
\email{$^1$ hrspksc@hkucc.hku.hk, $^2$ harko@hkucc.hku.hk}
\affil{$^{1,2}$ Department of Physics, The University of Hong Kong, Pokfulam Road, Hong Kong SAR, P.R.~China}

\begin{abstract}
We consider the bremsstrahlung surface photon emissivity of strange quark
stars, by sistematically taking into account the effects of the multiple
scatterings of the highly relativistic quarks in a dense medium (the
Landau-Pomeranchuk-Migdal effect). Due to interference between amplitudes of
nearby interactions the bremsstrahlung emissivity from strange star surface
is suppressed for frequencies smaller than a critical frequency. The range
of the suppressed frequencies is a function of the quark matter density at the
star's surface and of the QCD coupling constant. For temperatures much smaller than the
Fermi energy of the quarks the bremsstrahlung spectrum has the same temperature dependence
as the equilibrium black body radiation. Multiple collisions could reduce by an
order of magnitude the intensity of the bremsstrahlung radiation.
The effect of the thin electron layer at the surface of the
quark star on the bremsstrahlung spectrum is also analyzed in detail. It is shown
that absorption in the semi-degenerate electron gas can also significantly reduce
the intensity of the quark-quark bremsstrahlung radiation and, consequently, the surface emissivity.
Hence the combined effects of multiple collisions and absorption in the electron layer
could make the soft photon surface radiation of quark stars six orders of magnitude
smaller than the equilibrium black body radiation.
\end{abstract}

\keywords{radiation mechanisms - general: quark stars}

\section{Introduction}

The quark structure of the nucleons, suggested by quantum chromodynamics,
opens the possibility of a hadron-quark phase transition at high densities
and/or temperatures, as suggested by \citet{Wi84}. In the theories of strong
interaction quark bag models suppose that breaking of physical vacuum takes
place inside hadrons. If the hypothesis of the quark matter is true, then
some compact objects identified with neutron stars could actually be strange stars, built entirely of
strange matter \citep{Al86,Ha86}. For a review of
strange star properties, see \citet{Ch98}.

Most of the investigations of quark star properties have been done within
the framework of the so-called MIT bag model. Assuming that interactions of
quarks and gluons are sufficiently small, neglecting quark masses and
supposing that quarks are confined to the bag volume (in the case of a bare
strange star, the boundary of the bag coincides with stellar surface), the
energy density $\rho c^{2}$ and pressure $p$ of a quark-gluon plasma are
related, in the MIT bag model, by the equation of state (EOS) \citep{Ch98} 
\begin{equation}
p=\frac{(\rho -4B)c^{2}}{3},  \label{1}
\end{equation}
where $B$ is the difference between the energy density of the perturbative
and non-perturbative QCD vacuum (the bag constant). Equation (\ref{1}) is
essentially the equation of state of a gas of massless particles with
corrections due to the QCD trace anomaly and perturbative interactions.

More sophisticated investigations of quark-gluon interactions have shown
that Eq.~(\ref{1}) represents a limiting case of more general equations of
state. For example MIT bag models with massive strange quarks and lowest
order QCD interactions lead to some corrections terms in the equation of
state of quark matter. Models incorporating restoration of chiral quark
masses at high densities and giving absolutely stable strange matter can no
longer be accurately described by using Eq.~(\ref{1}). On the other hand in
models in which quark interaction is described by an interquark potential
originating from gluon exchange and by a density dependent scalar potential
which restores the chiral symmetry at high densities \citep{De98}, the
equation of state $P=P\left( \rho \right) $ can be well approximated by a
linear function in the energy density $\rho $ \citep{Go00}. \citet{Zd00} has
studied the linear approximation of the equation of state, obtaining all
parameters of the EOS as polynomial functions of strange quark mass, QCD
coupling constant and bag constant.

A complete description of static strange stars has been obtained based on
numerical integration of mass continuity and TOV (hydrostatic equilibrium)
equations for different values of the bag constant \citep{Wi84,Ha86}. Using numerical methods the maximum gravitational mass $M_{max}$
and the maximum radius $R_{max}$ of the strange star , have been obtained,
as a function of the bag constant, in the form \citep{Wi84,Ha86,Go,ChHa,HaCh02}
\begin{equation}
M_{max}=\frac {1.9638M_{\odot }}{\sqrt {B_{60}}},R_{max}=\frac {10.172\textrm{
km }}{\sqrt {B_{60}}},
\end{equation}
where $B_{60}\equiv B/(60$ MeV fm$^{-3})$.

There are several proposed mechanisms for the formation of quark stars.
Quark stars are expected to form during the collapse of the core of a
massive star after the supernova explosion as a result of a first or second
order phase transition, resulting in deconfined quark matter \citep{Da}. The
proto-neutron star core or the neutron star core is a favorable environment
for the conversion of ordinary matter to strange quark matter \citep{ChDa}.
Another possibility is that some neutron stars in low-mass X-ray binaries
can accrete sufficient mass to undergo a phase transition to become strange
stars \citep{Ch96}. This mechanism has also been proposed as a source of
radiation emission for cosmological $\gamma $-ray bursts \citep{Ch98a}.

Strange quark matter is filled with electromagnetic waves in thermodynamic
equilibrium with quarks. Photon emissivity is the basic parameter for
determining macroscopic properties of objects made of strange matter.
\citet{Al86} have shown that, because of very high
plasma frequency near the strange matter edge, photon emissivity of strange
matter is very low. For temperatures $T<<E_{p}/k$, where $E_{p}\approx 23$
MeV is the characteristic transverse plasmon cuttof energy, the equilibrium
photon emissivity of strange matter is negligible small, compared to the
black body one. The dispersion relation of the electromagnetic waves in
quark matter can be written as $\omega =\left( \omega
_{p}^{2}+k^{2}c^{2}\right) ^{1/2}$, where $\omega $ is the frequency of the
waves, $k$ is their wave number, $\omega _{p}=\left( 8\pi
e^{2}c^{2}n_{b}/3\mu \right) ^{1/2}$ is the plasma frequency, with $n_{b}$
the baryon number density of strange quark matter and $\mu \approx \hbar
c\left( \pi ^{2}n_{b}\right) ^{1/3}$ is the chemical potential of the quarks
(Usov 1998). Propagating modes exist only for $\omega >\omega _{p}$. Hence
the lower limit of the energy $E_{\gamma }$ of the electromagnetic quanta,
which are in equilibrium with quarks is given by $E_{\gamma }=\hbar \omega
>\hbar \omega _{p}\approx 18.5\left( n_{b}/n_{0}\right) ^{1/3}$ MeV, where $%
n_{0}=0.17$ fm$^{-3}$ is normal nuclear matter density. At the quark star
surface $n_{b}\approx \left( 1.5-2\right) n_{0}$ and the spectrum of
equilibrium photons is very hard, with $\hbar \omega \approx 20-25$ MeV \citep
{Chmaj}. The bremsstrahlung emissivity of quark matter has been estimated by 
\citet{Chmaj}. They have found that the surface radiation is about four
orders of magnitude weaker than the equilibrium black body radiation, even
they both have the same temperature dependence.

The structure of a realistic strange star is very complicated, but its basic
properties can be described as follows \citep{Al86}. Beta-equilibrated strange
quark - star matter consists of an approximately equal mixture of up $u$, down $d$
and strange $s$ quarks, with a slight deficit of the latter. The Fermi gas of $%
3A $ quarks constitutes a single color-singlet baryon, with baryon number $A$%
. This structure of the quarks leads to a net positive charge inside the
star. Since stars in their lowest energy state are supposed to be charge
neutral, electrons must balance the net positive quark charge in strange
matter stars. The electrons, being bounded to the quark matter by the
electromagnetic interaction and not by the strong force, are able to move
freely across the quark surface, but clearly cannot move to infinity because
of the electrostatic attraction of quarks. The electron distribution extends
up to $\sim 10^{3}$ fm above the quark surface.

The Coulomb barrier at the
quark surface of a hot strange star may also be a powerful source of $%
e^{+}e^{-}$ pairs, which are created in the extremely strong electric field of the
barrier. At surface temperatures of around $10^{11}$ K, the luminosity of
the outflowing plasma may be of the order $\sim 10^{51}$ ergs$^{-1}$ \citep
{Us98}. Moreover, as shown by \citet{PaUs02}, the thermal luminosity from the star surface, due to both
photon emission and $e^{+}e^{-}$ pair production may be, for about one day
for normal quark matter and for up to a hundred years for superconducting quark matter,
orders of magnitude higher than the Eddington limit.

It is the purpose of the present paper to reconsider the problem of the
photon emissivity, via bremsstrahlung radiation, of the quark stars.
Equilibrium radiation, transmitted through the surface, is dominant at
temperatures $T>2\times 10^{10}$ K, while below this temperature the
bremsstrahlung radiation from the surface layer prevails. Hence
bremsstrahlung is the main source of radiation for cold quark stars. This
electromagnetic radiation is generated in the dense medium at the stellar
surface in which extremely relativistic quarks move. In the 1950's \citet{LaPo53} predicted that the cross section for bremsstrahlung
from highly relativistic particles in dense media is suppressed, due to the
interference between amplitudes of nearby interactions. The suppression has
its roots in the uncertainty principle. The kinematics of the bremsstrahlung
requires that the longitudinal momentum transfer between the fixed and the
scattered particles must be small. On the other hand the uncertainty
principle requires that the interaction must occur over a large longitudinal
distance scale (formation zone). If the charged particle Coulomb scatters
while traversing this zone, the bremsstrahlung amplitude from before and
after the scattering can interfere, reducing the amplitude for photon
emission \citep{LaPo53,An95,Kl99}.
This effect has been recently confirmed experimentally for the bremsstrahlung
emission of electrons in different materials at a 5\% accuracy
level \citep{An95}. Since the surface of the quark star consists of a very
dense medium, with density of the order $4\times B\approx4\times 10^{14}$,
multiple collisions of the electromagnetically radiating particle lead to a significant suppression of the
quark-quark bremsstrahlung. By adopting a simple model for the elastic scattering
of quarks, we derive the spectrum of the classical bremsstrahlung radiation and the
emissivity of the quark matter, by considering the multiple scattering effects in the electromagnetic radiation 
of quark matter. For the range of densities existing at the surfaces of strange stars,
which are higher than the nuclear density, the Landau-Pomeranchuk effect could reduce
the bremsstrahlung emissivity of quark matter by an order of magnitude.

Photon emissivity of quark-gluon plasma (QGP), which is conjectured to be
formed in ultrarelativistic heavy ion collisions, has been extensively
investigated recently (for a recent review of direct photon emission from
QGP, including comparisons of theoretical predictions with experiments see
Peitzmann \& Thoma (2002)). Photons and dilepton pairs only interact
electromagnetically and their mean free paths are much larger than the size
of the QGP. Hence these electromagnetic probes leave the hot and dense QGP
without further scattering, thus providing important informations about the
early stages of the collision and the structure of the QGP. As sources of
direct photons one can consider quark annihilation, Compton scattering and
bremsstrahlung following the initial hard scattering of partons of the
nuclei, as well as thermal photons from the QGP and from hadronic
interactions in the hot hadronic gas after the hadronization of the plasma
(Peitzmann \& Thoma 2002). The first observation of direct photon production
in ultrarelativistic heavy ion collisions has been reported by the WA98
collaboration in $^{208}Pb+^{208}Pb$ collisions at $\sqrt{s}=158$ GeV at the
Super Proton Synchrotron at CERN (Aggarwal et al. 2000). The results display
a clear excess of direct photons above the expected background from hadronic
decays in the range of transverse momentum $p_{T}>1.5$ GeV$/c$ in the most
central collisions. These experimental findings provide a confirmation of the
feasibility of direct photons as reliable probes in heavy ion collisions and
may pave the way for the understanding of the formation and evolution of the
QGP.

The evaluation of the photo-emission rate from the QGP was initiated by
Kapusta et al. (1991) and by Baier et al. (1992). However, as shown by
Aurenche et al. (2000), these initial estimations were incomplete, since the
bremsstrahlung and inelastic pair annihilation processes contains collinear
enhancements which cause them to contribute at the same parametric order in
the coupling as the two to two processes even for large photon energies. But
these calculations are also incomplete, as they did not incorporate the
suppression of photon emission due to multiple scatterings during the photon
emission process, which limits the coherence length of the emitted radiation
(the Landau-Pomeranchuk-Migdal effect). The rate of photo-production by
bremsstrahlung and inelastic pair annihilation in a hot, equilibrated plasma
at zero chemical potential, fully including the LPM effect, was calculated,
to leading orders in both the electromagnetic and strong coupling constants,
by (Arnold, Moore \& Jaffe 2001, Arnold, Moore \& Jaffe 2002). The emission
of hard photons from the QGP plasma, using a model based on the
thermodynamics of QCD, and the determination of the initial temperature of
the expanding fireball has been considered recently by Renk (2003).

The radiation emitted by the quark layer at the surface of the quark star
must also cross the thin electron layer extending beyond the star's surface.
We also consider the effect of this layer on the outgoing radiation and show
that it can significantly reduce the intensity of the radiation coming from
the quark star.

The present paper is organized as follows. In Section II we discuss the
Landau-Pomeranchuk-Migdal effect for quark matter. The bremsstrahlung
emissivity of quark matter is considered in Section III. The effect of the
electron layer on the radiation outgoing from the quark star surface is
analyzed in Section IV. In Section V we discuss and conclude our results.
Throughout this paper we use units so that $c=\hbar =k_{B}=1$, with $k_{B}$
the Boltzmann constant.

\section{Landau-Pomeranchuk-Migdal effect in quark matter}

The dispersion relation shows that only photons with frequency $\omega
>\omega _{p}$ can propagate inside strange matter. Propagation of
electromagnetic waves of frequencies lower than $\omega _{p}$ is
exponentially damped, with the damping coefficient
$\beta $ depending on $\omega $ and $\omega _{p}$. For $\omega <<\omega _{p}$ the value
of $\beta $ is given by $\beta \approx 2\omega _{p}$ \cite{Ja75}. 
This means that photons with $\omega <\omega _{p}$
emitted in various processes inside strange matter are absorbed. The mean
free path of such a photon is of the order of
$\lambda \approx c/\beta \approx c/2\omega _{p}\approx 5$ fm \citep{Chmaj}.
Therefore, only
photons produced just below the surface with momenta pointing outwards can
leave strange matter.

In the following we shall consider the photon emissivity of quark matter, by
also taking into account the effect of multiple scatterings on the emitted
radiation. This effect is called the Landau-Pomeranchuk-Migdal (LPM) effect,
and was considered initially by \citet{LaPo53}, who
used classical electrodynamics to demonstrate bremsstrahlung suppression due
to multiple scattering: the suppression comes from the interference between
photons emitted by different elements of the charged particle trajectory. To
study the astrophysical implications of this effect on quark matter we shall
take into account the bremsstrahlung photons produced in the quark-quark
interactions, that is in the process $q_{1}+q_{2}\rightarrow q_{1^{\prime
}}+q_{2^{\prime }}+\gamma $.

The LPM effect can be discussed in a simple qualitative way as follows.
Let's consider a bremsstrahlung photon of energy $\omega $ produced by an
ultrarelativistic quark of energy $E$ and mass $m_{eff}$, where $\omega <<E$%
. In this kinematic regime the average angle $\theta _{\gamma }$ between the
incident quark and the produced photon is small, $\theta _{\gamma }\approx
m_{eff}/E$. We assume that the angle between the incident and scattered
quark is also small. Moreover, we shall assume that all quarks have the same
density dependent effective mass $m_{eff}\approx \mu \sqrt{4\alpha _{s}/3\pi 
}\approx \sqrt{4\alpha _{s}/3\pi }\left( \pi ^{2}n_{b}\right) ^{1/3}$ ,
where $\alpha _{s}$ is the QCD coupling constant and $n_{b}=\left(
n_{u}+n_{d}+n_{s}\right) /3$, with $n_{j},j=u,d,s,...$ the quark particle
number densities. Neglecting the scattering and quark-photon angles, it
follows that the longitudinal momentum transfer $p_{\parallel }$ to the
static quark is $p_{\parallel }\approx \omega /2\gamma ^{2}$, where $\gamma
=E/m_{eff}$ \citep{Kl99}. The uncertainty principle then requires that the
spatial position of the bremsstrahlung process has a longitudinal
uncertainty of $L_{\parallel }\sim 1/p_{\parallel }\approx 2\gamma
^{2}/\omega $. In alternate language, the quark and photon slowly split
apart over the distance $L_{\parallel }$. In a sufficiently dense medium the
quark mean free path is much less than $L_{\parallel }$; in the LPM effect
the relevant interaction is multiple scattering.

Bremsstrahlung is suppressed when the mean square multiple scattering angle over the distance $%
L_{\parallel }$, $\theta _{ms}^{2}=$ $\left( E_{s}/E\right) ^{2}\left(
L_{\parallel }/X_{0}\right) $, where $E_{s}=m_{eff}\sqrt{4\pi /\alpha }$ and 
$X_{0}=\left[ 4n\alpha r_{q}^{2}Z^{2}\ln \left( 184Z^{-1/3}\right) \right]
^{-1}$, with $r_{q}=e^{2}/m_{eff}$, is the radiation length, is greater than
or equal to $\theta _{\gamma }^{2}$, $\theta _{ms}^{2}\geq \theta _{ms}^{2}$
\citep{An95,Kl99}. Therefore the bremsstrahlung differential
cross section for the production of a photon is suppressed when $\omega
<\omega _{LPM}=E^{2}/E_{LPM}$, where $E_{LPM}=m_{eff}^{2}X_{0}\alpha /8\pi $
\citep{An95,Kl99}. By assuming that the temperature of the star $%
T<<p_{F}$, with $p_{F}$ the Fermi momentum, the Fermi distribution factors
force all the quarks to have $E\approx \mu $. For an extreme relativistic
quark gas $E\approx p_{F}=\left( 6\pi ^{2}n_{b}/g\right) ^{1/3}$, where $g$
is the statistical weight for quark matter. Therefore the critical LPM
frequency for bremsstrahlung photon emission in quark matter is given by 
\begin{equation}
\omega _{LPM}=\frac{2^{1/3}3^{4/3}\pi ^{5/3}g^{2/3}e^{4}Z^{2}\ln \left(
184Z^{-1/3}\right) }{\alpha _{s}^{2}}n_{b}^{1/3}=an_{b}^{1/3},
\end{equation}
where $a=2^{1/3}3^{4/3}\pi ^{5/3}g^{2/3}e^{4}Z^{2}\ln \left(
184Z^{-1/3}\right) /\alpha _{s}^{2}$. For strange quark matter $g=6$.

Emission of the bremsstrahlung radiation with frequency smaller than $\omega
_{LPM}$ is suppressed, due to the effect of multiple scattering. As
expected, the LPM frequency increases with the density of the strange
matter. The variation of $\omega _{LPM}$ as a function of the particle
number density is represented, for different values of the strong coupling
constant $\alpha _{s}$, in Fig.~1.

The most important parameter characterizing the LPM effect is the average
time between two collisions, denoted $\tau $. It can be calculated from $%
\tau ^{-1}=n\sigma _{0}v$, where $n$ is the density of the medium which
depends on its composition and on temperature \citep{ClGoRe93}. $\sigma _{0}$
denotes the cross section of a test particle having velocity $v$ and
interacting with the particles of the medium. Thus generally $\tau $ is a
function of the temperature and density of the medium and of the momentum of
the test particle. The value of $\tau $ is essential for the importance of
the LPM effect: if $\tau ^{-1}$ is zero ($\tau \rightarrow \infty $), the
effect is absent \citep{ClGoRe93}.

The starting point for the rigorous implementation of the LPM effect is the
following textbook equation for the energy radiated per unit momentum
by the charged particle \citep{Ja75}, 
\begin{equation}
\frac{dI}{d^{3}k}=\frac{\alpha }{\left( 2\pi \right) ^{2}}%
\sum_{l=u,d,s}\left( \frac{q_{l}}{e}\right) ^{2}\left| \int_{-\infty
}^{+\infty }e^{i\left[ \omega t-\vec{k}\cdot \vec{r}_{l}\left( t\right) %
\right] }\vec{n}\times \vec{v}_{l}dt\right| ^{2},  \label{1a}
\end{equation}
where $\vec{r}_{l}\left( t\right) $ describes the trajectory of the charged
particle, $\vec{v}_{l}\left( t\right) $ is its velocity, $\vec{n}=\vec{k}%
/\left| \vec{k}\right| $, $\alpha $ is the fine structure constant and $%
q_{l} $ is the charge of $\left( u,d,s\right) $ quarks. If only one
collision occurs and if the velocity is a constant ($\vec{v}=\vec{v}_{1}$)
before and after the collision ($\vec{v}=\vec{v}_{2}$), then the radiated energy
is given by \citep{Ja75} 
\begin{eqnarray}\label{single}
\frac{dI}{d^{3}k} &=&\frac{\alpha }{\left( 2\pi \right) ^{2}}%
\sum_{l=u,d,s}\left( \frac{q_{l}}{e}\right) ^{2}\left| \int_{-\infty
}^{t_{1}}e^{i\left( \omega -\vec{k}\cdot \vec{v}_{1l}\right) t}\vec{n}\times 
\vec{v}_{1l}dt+\int_{-\infty }^{t_{1}}e^{i\left( \omega -\vec{k}\cdot \vec{v}%
_{2l}\right) t}\vec{n}\times \vec{v}_{2l}dt\right| ^{2}  \nonumber
\label{2a} \\
&=&\frac{\alpha }{\left( 2\pi \right) ^{2}}\sum_{l=u,d,s}\left( \frac{q_{l}}{%
e}\right) ^{2}\left| \frac{\vec{n}\times \vec{v}_{1l}}{\omega -\vec{k}\cdot 
\vec{v}_{1l}}+\frac{\vec{n}\times \vec{v}_{2l}}{\omega -\vec{k}\cdot \vec{v}%
_{2l}}\right| ^{2},
\end{eqnarray}
which is the standard textbook form for the calculation of the intensity.

If, however, a series of collisions occurs, one has to change Eqs.~(\ref{1a}%
) and (\ref{2a}) to a sum over all pieces of the trajectory \citep{ClGoRe93}, 
\begin{eqnarray}
\frac{dI}{d^{3}k}&=&\frac{\alpha }{\left( 2\pi \right) ^{2}}%
\sum_{l=u,d,s}\left( \frac{q_{l}}{e}\right) ^{2}\\\nonumber 
&&\left| \int_{-\infty
}^{t_{1}}e^{i\left[ \omega t-\vec{k}\cdot \vec{r}_{l}\left( t\right) \right]
}\vec{n}\times \vec{v}_{1l}dt+\int_{t_{1}}^{t_{2}}e^{i\left[ \omega t-\vec{k}%
\cdot \vec{r}_{l}\left( t\right) \right] }\vec{n}\times \vec{v}%
_{2l}dt+...+\int_{t_{n-1}}^{t_{n}}e^{i\left[ \omega t-\vec{k}\cdot \vec{r}%
_{l}\left( t\right) \right] }\vec{n}\times \vec{v}_{nl}dt+....\right| ^{2}
\end{eqnarray}

Assuming that the velocity is a constant between two collisions leads to 
\begin{equation}
\frac{dI}{d^{3}k}=\frac{\alpha }{\left( 2\pi \right) ^{2}}%
\sum_{l=u,d,s}\left( \frac{q_{l}}{e}\right) ^{2}\left|
\sum_{j=1}^{N}e^{i\left( \omega t_{j-1l}-\vec{k}\cdot \vec{r}_{j-1l}\right)
}\left( 1-e^{-i\left( \omega -\vec{k}\cdot \vec{v}_{jl}\right) \left(
t_{j}-t_{j-1}\right) }\right) \right| ^{2},
\end{equation}
where $N$ is the total number of collisions. The calculation of the square
gives 
\begin{eqnarray}
\frac{dI}{d^{3}k} &=&\frac{\alpha }{\left( 2\pi \right) ^{2}}%
\sum_{l=u,d,s}\left( \frac{q_{l}}{e}\right) ^{2}\sum_{j=1}^{N}\frac{\vec{v}%
_{jl}^{2}-\left( \vec{n}\cdot \vec{v}_{jl}\right) ^{2}}{\left( \omega -\vec{k%
}\cdot \vec{v}_{jl}\right) ^{2}}\left[ 2-e^{i\xi _{j}\left( \omega -\vec{k}%
\cdot \vec{v}_{jl}\right) }-e^{-i\xi _{j}\left( \omega -\vec{k}\cdot \vec{v}%
_{jl}\right) }\right] +  \nonumber \\
&&2\frac{\alpha }{\left( 2\pi \right) ^{2}}\sum_{m=u,d,s}\left( \frac{q_{m}}{%
e}\right) ^{2}\textrm{Re}\sum_{j>l}^{N}\frac{\vec{v}_{jm}\cdot \vec{v}%
_{lm}-\left( \vec{n}\cdot \vec{v}_{jm}\right) \left( \vec{n}\cdot \vec{v}%
_{lm}\right) }{\left( \omega -\vec{k}\cdot \vec{v}_{jm}\right) \left( \omega
-\vec{k}\cdot \vec{v}_{lm}\right) }\exp \left[ i\sum_{i=l}^{j-1}\xi
_{i}\left( \omega -\vec{k}\cdot \vec{v}_{im}\right) \right]\nonumber \\
&&\left(
1-e^{-i\left( \omega -\vec{k}\cdot \vec{v}_{jm}\right) \xi _{j}}\right)
\times  \left( 1-e^{-i\left( \omega -\vec{k}\cdot \vec{v}_{lm}\right) \xi
_{l}}\right) ,
\end{eqnarray}
where we denoted $\xi _{j}=t_{j}-t_{j-1}$.

As the quark gas is taken to be in thermal equilibrium, many possible
velocities can result in between two collisions. For this reason the
non-diagonal terms in the above equation will give zero contribution after
one averages over all velocities, providing of course that no correlation
exists between the velocity before and after the collision. Only the
diagonal terms remain in this case. The time between two successive
collisions is given by $\xi $. Taking an average over the time between two
collisions can be obtained using the following distribution \citep{ClGoRe93} 
\begin{equation}
\frac{dW}{d\xi }=\frac{1}{\tau }e^{-\frac{\xi }{\tau }}.
\end{equation}

Then the electromagnetic energy radiated away by the quarks is given by 
\begin{equation}
\frac{dI}{d^{3}k}=\frac{\alpha }{\left( 2\pi \right) ^{2}}%
\sum_{l=u,d,s}\left( \frac{q_{l}}{e}\right) ^{2}\frac{1}{\tau }%
\int_{0}^{\infty }e^{-\xi \tau ^{-1}}\frac{v_{l}^{2}-\left( \vec{n}\cdot 
\vec{v}_{l}\right) ^{2}}{\left( \omega -\vec{k}\cdot \vec{v}_{l}\right) ^{2}}%
\left[ 2-e^{i\left( \omega -\vec{k}\cdot \vec{v}_{l}\right) \xi
}-e^{-i\left( \omega -\vec{k}\cdot \vec{v}_{l}\right) \xi }\right] d\xi ,
\end{equation}
where we have introduced the average velocities by means of the definitions $%
\vec{v}=\left\langle \vec{v}\right\rangle =\sum_{j=1}^{N}\vec{v}_{j}/N$, $%
\vec{v}^{2}=\left\langle \vec{v}^{2}\right\rangle =\sum_{j=1}^{N}\vec{v}%
_{j}^{2}/N$ and we also denoted $v=\left| \left\langle \vec{v}\right\rangle
\right| $.

The averaging over the time between two collisions can be done with the use
of the mathematical identity $a\int_{0}^{\infty }\left( 1-\cos b\xi \right)
e^{-a\xi }d\xi =b^{2}/\left( a^{2}+b^{2}\right) $. Therefore we obtain for
the intensity of the radiation emitted by a quark moving in a dense medium
the expression  
\begin{equation}
\frac{dI}{d^{3}k}=\frac{\alpha }{2\pi ^{2}}\sum_{l=u,d,s}\left( \frac{q_{l}}{%
e}\right) ^{2}\frac{v_{l}^{2}-\left( \vec{n}\cdot \vec{v}_{l}\right) ^{2}}{%
\tau ^{-2}+\left( \omega -\vec{k}\cdot \vec{v}\right) ^{2}}.
\end{equation}

Since $d^{3}k=\omega ^{2}d\omega d\Omega $, the intensity can also be
written as 
\begin{equation}
dI=\frac{\alpha }{2\pi ^{2}}\sum_{l=u,d,s}\left( \frac{q_{l}}{e}\right)
^{2}v_{l}^{2}\frac{1-\cos ^{2}\theta }{\tau ^{-2}+\left( \omega -kv_{l}\cos
\theta \right) ^{2}}\omega ^{2}d\omega d\Omega .
\end{equation}

The angular integration is standard, and finally we obtain for the intensity
of the radiation emitted by the quark moving in a dense medium the
expression \citep{ClGoRe93}
\begin{equation}
dI=\frac{\alpha }{\pi }\sum_{l=u,d,s}\left( \frac{q_{l}}{e}\right) ^{2}\frac{%
1}{v_{l}}\Phi \left( \omega ,v_{l},\tau ^{-1}\right) \frac{d\omega }{\omega }%
,  \label{int}
\end{equation}
where the function $\Phi \left( \omega ,v,\tau ^{-1}\right) $ is given by 
\begin{eqnarray}
\Phi \left( \omega ,v_{l},\tau ^{-1}\right) &=&\frac{1-\tau ^{2}\omega
^{2}\left( 1-v_{l}^{2}\right) }{\tau }\left[ \arctan \tau \omega \left(
1+v_{l}\right) -\arctan \tau \omega \left( 1-v_{l}\right) \right] \nonumber \\
&&+\omega
\ln \frac{1+\omega ^{2}\tau ^{2}\left( 1+v_{l}\right) ^{2}}{1+\omega
^{2}\tau ^{2}\left( 1-v_{l}\right) ^{2}}-2\omega v_{l}.
\end{eqnarray}

In order to compute the numerical values of the intensity we need to know
the mean velocities of the quarks randomly moving in a dense medium. The
multiple scattering of the quarks change the quark path and affect the
emission of the radiation. If $\vec{v}(0)$ is the initial velocity of the
quark at the moment $t=0$, then at the moment $t$ its velocity will be $\vec{%
v}\left( t\right) =\vec{v}_{z}(t)+\vec{v}_{\perp }(t)=\vec{v}\left( 0\right)
\left( 1-\theta _{ms}^{2}(t)/2\right) +\left| \vec{v}\left( 0\right) \right|
\theta _{ms}(t)\vec{\Theta}$, where $\vec{\Theta}$ is a unit vector
perpendicular to the initial direction of motion and $\theta _{ms}(t)$ is
the electron multiple scattering in the time interval $0$ to $t$ \citep{Kl99}. \citet{LaPo53} took $\theta _{ms}^{2}\approx
\left\langle \theta _{ms}^{2}\right\rangle $. Therefore, in this
approximation we obtain $v=v\left( 0\right) \sqrt{1+\left\langle \theta
_{ms}^{2}\right\rangle ^{2}/4}$. Over a distance $\vec{d}=\vec{v}_{z}t$ the
multiple scattering is given by $\left\langle \theta _{ms}^{2}\right\rangle
=\left( E_{s}/E\right) ^{2}\left( d/X_{0}\right) $. For $d$ we shall assume
that it is the suppression distance, that is, the distance over the
interaction actually spread: $d=l_{f}$. In the most general case, with
quantum corrections taken into account, $l_{f}$ is given as a root of a
quadratic equation: $l_{f}=l_{f0}\left(
1+E_{s}^{2}l_{f}/2m_{eff}^{2}X_{0}\right) ^{-1}$, where $l_{f0}=2E^{2}/%
\omega m_{eff}^{2}$ is the classical suppression length \citep{Kl99}.

In the case of the standard approach to the bremsstrahlung radiation of
charged particles, one assumes that the energy is emitted in a single
collision. Then, from Eq.~(\ref{single}), by taking $\vec{v}_{2}=0$ and $\vec{%
v}_{1}=\vec{v}$, it immediately follows that the single collision spectral
distribution of the bremsstrahlung radiation $dI_{sc}$ is given by \citep{LaLi75}
\begin{equation}\label{bremst}
dI_{sc}=\frac{\alpha }{\pi }\sum_{l=u,d,s}\left( \frac{q_{l}}{e}\right)
^{2}\left( \frac{1}{v_{l}}\ln \frac{1+v_{l}}{1-v_{l}}-2\right) d\omega .
\end{equation}

The total intensity is proportional to the frequency of the emitted
radiation and is independent on the medium in which the motion of the
particle takes place. The same result can be derived by taking the limit $\tau
\rightarrow \infty $ in Eq.~(\ref{int}):
\begin{equation}
\frac{dI_{sc}}{d\omega }=\lim_{\tau \rightarrow \infty }\frac{dI}{d\omega }.
\end{equation}

In the opposite limit of very high density, corresponding to $\tau
\rightarrow 0$, from Eq.~(\ref{int}) we obtain
\begin{equation}
\lim_{\tau \rightarrow 0}\frac{dI}{d\omega }=0,
\end{equation}
and $\lim_{\tau \rightarrow 0}I=0$. Therefore the electromagnetic radiation
from a quark moving in a very dense medium is completely suppressed.

\section{Bremsstrahlung emissivity of strange stars}

To calculate the bremsstrahlung emissivity a sensible model for quark
interactions has to be used. We adopt the basic assumption that the details
of the interaction should be relatively unimportant, as long as the overall
strength is correct and the relevant symmetries are respected. Hence we use
the well-known Nambu-Jona-Lasinio (NJL) model in its $SU(2)$ version in our
calculations \citep{HaKu94}. It has the same symmetries as QCD
and describes an effective pointlike interaction with a constant coupling
strength. Usually the NJL model and its extensions are used in mean field
calculations. As a consequence of adopting this model all the aspects
related to color superconductivity are neglected.

Let $d\sigma _{0}$ be the cross-section for a given process of scattering of
charged particles, which may be accompanied by the emission of a certain
number of photons. For example, $d\sigma _{0}$ could refer to the scattering
of a quark by an other quark, with the possible emission of hard photons.
Together with this process one could consider another process which differs
from it only in that one extra photon is emitted. In this case the total
cross-section $d\sigma $ can be represented as a product of two independent
factors, the cross-section $d\sigma _{0}$ and the probability $dW_{\gamma }$
of emission of a single photon in the collision \citep{La}. The emission of a
soft photon is a quasi-classical process. The probability of emission is the
same as the classically calculated number of quanta emitted in the
collision, that is the same as the classical intensity (total energy) of
emission $dI$, divided by the frequency of the radiation $\omega $ \citep{La}%
. Hence as a first step in obtaining the bremsstrahlung emissivity of quark
matter we have to calculate the classical radiation intensity emitted by a
quark moving in a dense medium in which many inter-particle collisions occur.

The probability $dW_{\gamma }$ of emitting a photon of energy $\omega $ by
any of the scattered quarks is $dW_{\gamma }=dI/\omega $. Therefore the
total cross section for the emission of soft bremsstrahlung photons is given
by \citep{La} 
\begin{equation}
d\sigma =d\sigma _{0}\frac{dI}{\omega }.  \label{cross}
\end{equation}

To calculate the total cross section for a photon emission we need to know $%
d\sigma _{0}$, describing the scattering of quarks. Within the two-flavor
NJL model the elementary quark-quark cross sections have been calculated by
\citet{ZhHuKlNe95}.The calculations have been done
nonperturbatively in the coupling constant, using the so-called $1/N_{c}$
expansion, where $N_{c}$ is the number of colors. In the first order of the $%
1/N_{c}$ expansion and for temperature $T<T_{c}$, where $T_{c}=0.19$ GeV,
the quark-quark scattering cross-sections have a remarkably simple
dependence on the collision energy $s$, given by $\sigma _{0}(s)\sim 1/s$ \citep{ZhHuKlNe95}.
The cross section decreases with increasing energy. For a more general
parametrization of the quark-quark scattering cross section, with the
inclusion of the double scattering terms see also \citet{Ma97}.

We assume that the quarks form a degenerate Fermi gas, with the particle
number density $n$ given by 
\begin{equation}
n=\frac{g}{2\pi ^{2}}\int_{0}^{\infty }f\left[ E(p)-\mu \right] p^{2}dp=%
\frac{g}{2\pi ^{2}}\int_{0}^{\infty }\frac{p^{2}dp}{\exp \left[ \frac{%
E(p)-\mu }{T}\right] +1},
\end{equation}
where $E=\left( p^{2}+m^{2}\right) ^{1/2}$ is the quark kinetic energy and $%
f(\left[ E(p)-\mu \right] =1/\left( \exp \left[ \frac{E(p)-\mu }{T}\right]
+1\right) $ is the Fermi-Dirac distribution function \citep{Ch68}.

In order to estimate the importance of the LPM effect on the bremsstrahlung
emissivity of quark matter we have to estimate $\tau $, the average time
between two collisions. In the case of the interaction of two quarks,
denoted 1 and 2, the average time can be obtained from \citep{HaPr94}
\begin{equation}
\tau ^{-1}=\int ds\frac{d^{3}p_{2}}{\left( 2\pi \right) ^{3}}f\left(
p_{2}\right) \sigma _{12}\left( s\right) v_{rel}\delta \left[ s-\left(
p_{1}+p_{2}\right) ^{2}\right] ,
\end{equation}
where 
\begin{equation}
v_{rel}=\left| \vec{v}_{1}-\vec{v}_{2}\right| =\frac{\sqrt{\left( p_{1}\cdot
p_{2}\right) ^{2}-m_{1}^{2}m_{2}^{2}}}{E_{1}E_{2}}=\frac{\sqrt{s\left(
s-4m_{eff}^{2}\right) }}{2E_{1}E_{2}}.
\end{equation}

The kinematic invariant $s$ is given by $s=2\left(
m_{eff}^{2}+E_{1}E_{2}-p_{1}p_{2}\cos \theta \right) $ \citep{ClGoRe93}. By assuming that
quarks are ultrarelativistic, with $E=p$, the average collision time can be
approximated by the following analytic representation: 
\begin{equation}
\tau ^{-1}\approx -\frac{1}{8\pi ^{8/3}}n_{b}^{-1/3}T^{2}Li_{2}\left[ -\exp
\left( \frac{\left( \pi ^{2}n_{b}\right) ^{1/3}}{T}\right) \right] ,
\end{equation}
where $Li_{2}(z)$ is the polylogarithm function defined as $%
Li_{n}(z)=\sum_{k=1}^{\infty }z^{k}/k^{n}$.

The variation of the collision time as a function of the quark
number density is presented, for different values of the temperature, in
Fig.~2.

In the framework of our model and for the range of temperatures considered,
the collision time is almost independent of $T$. As expected, $\tau $
decreases with increasing density. To calculate $\tau $ we have assumed
again that the test quarks, moving in the dense medium, have energies of the
same order as the Fermi energy, $E_{1}\approx E_{F}\approx p_{F}\approx
\left( \pi ^{2}n_{b}\right) ^{1/3}$. A rough estimate of the mean collision
time can be also obtained from the relation $\tau =1/n\sigma _{0}v$. By
assuming that the cross section of quark-quark interactions is around $%
\sigma _{0}\approx 1$ mb \citep{Ma97} and the quarks are travelling with
the speed of light, we obtain $\tau =41.6$ fm $\left( =12.48\times 10^{-23}%
\textrm{ s}\right) $ for $n_{b}=0.24$ fm$^{-3}$ and $\tau =25$ fm $\left(
=7.5\times 10^{-23}\textrm{ s}\right) $ for $n_{b}=0.4$ fm$^{-3}$.

The knowledge of $\tau $ allows the calculation of the spectrum of the
photons emitted via the bremsstrahlung mechanism from the quark star
surface. The applicability of the formulas for the spectral distribution and
intensity of the bremsstrahlung radiation we have derived in the previous
Section is limited by the quantum condition that $\omega $ be small as
compared with the total kinetic energy of the quark, which is of the order
of $E_{F}\approx \left( \pi ^{2}n_{b}\right) ^{1/3}$: $\omega <<\left( \pi
^{2}n_{b}\right) ^{1/3}$. Since at the star's surface the kinetic energy of
the quarks is of the order of $300$ MeV, the radiation formulae are valid
only for frequencies $\omega <<300$ MeV. In order to find the maximum
frequency $\omega _{\max }$ of the radiation in the ultrarelativistic case
we assume that it is of the order $\omega _{\max }\sim 1/\tau \left(
1-v^{2}\right) $ , that is it is inversely proportional to the duration of
the collision and to the square of the energy of the radiating particle
\citep{LaLi75}. If we assume that quarks have energies near the
Fermi energy, then the corresponding velocities are the Fermi velocities, 
\begin{equation}
v_{F}=\frac{\sqrt{3\pi /4\alpha _{s}}}{\sqrt{1+3\pi /4\alpha _{s}}},
\end{equation}
which are of the order of $v_{F}=0.92$ for $\alpha _{s}=0.4$ and $v_{F}=0.85$
for $\alpha _{s}=0.9$. Hence the maximum allowable frequency up to which the
formalism can be applied is given by
\begin{equation}
\omega _{\max }\sim \frac{1+3\pi /4\alpha _{s}}{\tau }\sim n_{b}\sigma _{0}%
\sqrt{3\pi /4\alpha _{s}\left( 1+3\pi /4\alpha _{s}\right) }.
\end{equation}

For a quark-quark interaction cross sections of the order of $1$ mb and for $%
\alpha _{s}=0.4$ we obtain $\omega _{\max }\sim 30$ MeV. In the following we
shall restrict our study only to quark bremsstrahlung frequencies satisfying
the condition $\omega <\omega _{\max }\approx 30$ MeV.   

The frequency distribution of the radiation emitted by a single
quark moving in the dense matter at the surface of the quark star is
presented in Fig.~3.

The total intensity $I$ of the bremsstrahlung radiation emitted by a single
quark, 
\begin{equation}
I=\frac{\alpha }{\pi }\sum_{l=u,d,s}\left( \frac{q_{l}}{e}\right)
^{2}\int_{\omega _{LPM}}^{\omega _{\max }}\frac{1}{v_{l}}\Phi \left( \omega
,v_{l},\tau ^{-1}\right) \frac{d\omega }{\omega },
\end{equation}
obtained by integrating Eq.~(\ref{int}) over all possible frequencies is
represented, as a function of the collision time $\tau $, in Fig.~4.

For the same range of frequencies the calculation of the bremsstrahlung intensity by using
Eq.~(\ref{bremst}), which does not take into account the effects of the multiple collisions, gives
$I\sim 0.11$ MeV. By taking into account that in the high density quark-gluon plasma the collision time is finite
and of the order of $\tau \sim 10$ fm leads to an intensity of the order $I\sim 0.03$ MeV.
Hence the LPM effect could reduce the intensity of the photon bremsstrahlung of quarks
at the surface of the star by a factor of $3$ to $5$ for $n_{b}\in \left( 0.24,0.4\right) $ fm$^{-3}$.
For collision times of the order of $\tau \sim 5$ fm, which, due to the incertitudes
in the quark-quark cross-sections, are also possible, the decrease in the intensity of the
bremsstrahlung is of an order of magnitude. 

The variation of the intensity of the electromagnetic radiation is represented, as a function of the quark density,
and for different values of the temperature, in Fig.~5

The intensity of the radiation emitted by a single particle is decreasing with
increasing density, due to the effects of the multiple scatterings of the
quarks. The number of photons is proportional to the number of collisions,
but if these happen too often, due to the high density of the medium, the
photons emitted at different points of the trajectory start to interfere
with each other and the intensity of the radiation is accordingly reduced. On
the other hand $I$ increases with the temperature of the dense quark matter.

If the quarks are extremely relativistic, with $v_{l}\rightarrow 1,l=u,d,s$,
the intensity of the total radiation emitted by a single quark can be
expressed in the following closed form
\begin{eqnarray}
\frac{\pi }{\alpha }I\left( n_{b},T\right)  &\approx &\frac{1+3\pi /4\alpha
_{s}}{\tau }\ln \left[ \frac{1+4\left( 1+3\pi /4\alpha _{s}\right) ^{2}}{%
\left( 1+4\tau ^{2}a^{2}n_{b}^{2/3}\right) ^{an_{b}^{1/3}}}\right] +  \nonumber
\\
&&\frac{1}{\tau }\left[ \arctan 2\left( 1+3\pi /4\alpha _{s}\right) -\arctan
2\tau an_{b}^{1/3}\right] - \nonumber\\
&&4\left( \frac{1+3\pi /4\alpha _{s}}{\tau }-an_{b}^{1/3}\right) +\frac{1}{%
2\tau }\left[ D\left( 1+3\pi /4\alpha _{s}\right) -D\left( \tau
an_{b}^{1/3}\right) \right] ,
\end{eqnarray}
where the function $D(x)$ is given by $D(x)=i\left[ Li_{2}\left( -2ix\right)
-Li_{2}\left( 2ix\right) \right] $. $D(x)$ is a real function for all $x$,
and its power series expansion is given by $%
D(x)=4x-16x^{3}/9+64x^{5}/25-256x^{7}/49+O\left( x^{8}\right) $. The
approximation $v\approx 1$ is appropriate for quarks having energies near
the Fermi energy and moving with velocities $v(0)=v_{F}$, which are of the order of $1$.

The emission of a photon in the quark matter is the result of the scattering
of a quark from a state $\vec{p}_{1}$ to a state $\vec{p}_{1}^{\prime }$ by
collision with another particle of momentum $\vec{p}_{2}$. The rate of
collisions in the quark-gluon  plasma in which a soft photon of energy $\omega <<E$
is emitted can be defined as \citep{Ch68} 
\begin{equation}
dN=g_{1}g_{2}f_{1}\left( \vec{p}_{1}\right) f_{2}\left( \vec{p}_{2}\right)
\sigma \left( \theta ,\phi \right) \left| \vec{v}_{1}-\vec{v}_{2}\right| %
\left[ 1-f_{1}\left( \vec{p}_{1}^{\prime }\right) \right] \left[
1-f_{2}\left( \vec{p}_{2}^{\prime }\right) \right] \prod_{i=1,2,1^{\prime
},2^{\prime }}d^{3}\vec{p}_{i}d\Omega \left( \theta ,\phi \right) ,
\end{equation}
where $d\Omega \left( \theta ,\phi \right) $ is the solid angle element in
the direction $\left( \theta ,\phi \right) $, $g_{1},g_{2}$ are the
statistical weights and $\vec{v}_{1}$, $\vec{v}_{2}$ are the velocities of
the interacting particles. The bremsstrahlung emissivity per unit volume of
the quark matter is given by 
\begin{equation}
\epsilon _{Br}=\frac{dE_{Br}}{dtdV}=\int \omega dN.
\end{equation}

Withe the use of the explicit form of the cross-section for the photon
emission, given by Eq.~(\ref{cross}), the photon emissivity can be written
in the following form: 
\begin{equation}\label{arbr}
\epsilon _{Br}=\chi _{Br}\left( n_{b},T\right) T^{4},
\end{equation}
where $\chi _{Br}$ is a function of the quark density and of the
temperature. 

The bremsstrahlung energy flux from the star, $F_{Br}$ is given by the
emissivity of a thin surface layer, with thickness of the order of the
photon mean free path $\lambda $ \citep{Chmaj}. Taking
into account only photons emitted outwards and produced not deeper than $\lambda $,
the energy flux from the surface of the quark star is given by 
\begin{equation}
F_{Br}=\sigma _{Br}\left( n_{b},T\right) T^{4},
\end{equation}
where $\sigma _{Br}\left( n_{b},T\right) =\chi _{Br}\lambda /\pi $. Since the
temperature dependence of $\sigma _{Br}$ is not significant, it follows that
the quark bremsstrahlung spectrum has the same temperature dependence as the
black body radiation.   

For arbitrary temperatures the coefficient of the bremsstrahlung emissivity can be represented, in an
approximate analytical form, by
\begin{equation}
\sigma _{Br}\left( n_{b},T\right) \approx \frac{g^{2}}{\left( 2\pi \right)
^{3}}Li_{2}^{2}\left[ -\exp \left( \frac{\left( \pi ^{2}n_{b}\right) ^{1/3}}{%
T}\right) \right] I\left( n_{b},T\right) \lambda .
\end{equation}

The calculation of $\epsilon _{Br}$ requires, in the general case, the
estimation of integrals of the form $F=\int_{0}^{\infty }pdp/\left[ \left(
\exp \left( p-\mu \right) /T\right) +1\right] $, which can be expressed
usually in terms of the polylogarithm function. In the limit of small
temperatures, with $T<<E_{F}=\left( \pi ^{2}n_{b}\right) ^{1/3}$, the
integral can be evaluated by means of the substitution $z=\left( p-\mu
\right) /T$. For small $T$ one can extend the lower limit of integration to $%
-\infty $, so that the integral becomes $F=\int_{-\infty }^{+\infty }T\left(
Tz+\mu \right) dz/\left[ \exp (z)+1\right] =\left( \pi ^{2}/6\right) T^{2}$.
Therefore in the limit of small temperatures $\epsilon _{Br}=\sigma
_{Br}\left( n_{b},T\right) T^{4}$, but there is no explicit temperature
dependence of the coefficient $\sigma _{Br}$:
\begin{equation}
\sigma _{Br}\left( n_{b},T\right) \approx \frac{g^{2}\pi }{576}I\left(
n_{b},T\right) \lambda .
\end{equation}

In this case the only $T$-dependence of $\sigma _{Br}$ is via the intensity
of the radiation.

The variation of the ratio of the bremsstrahlung emissivity coefficient and
the Stefan-Boltzmann constant $\sigma _{SB}$ is represented, as a function
of the mean collision time $\tau $, and in the limit of small temperatures,
in Fig.~6. We have assumed that the thickness of the emitting layer is $\lambda =5$
fm.

In the limit of large collisions times $\sigma _{Br}/\sigma _{SB}\sim 10^{-4}
$, and thus we recover the single-collision results of \citet{Chmaj}. For
collision times specific to the quark-gluon plasma at the surface of the
quark stars, of the order of $10$ fm, the ratio $\sigma _{Br}/\sigma
_{SB}\sim 10^{-5}$. Therefore, taking into account multiple collisions in
the high density plasma, reduces with an order of magnitude the value of the
bremsstrahlung coefficient $\sigma _{Br}$, making the soft photon radiation
from the surface of the quark star about $5$ orders of magnitude weaker than
the equilibrium black body radiation. 

\section{Effect of the electron layer on the photon emissivity of quark stars%
}

The radiation emitted by the thin quark surface layer of the strange star
must propagate through the electron layer formed due to the electrostatic
attraction of the quarks. In the followings we shall consider the effect of
the electron layer on the electromagnetic radiation emitted, via the
bremsstrahlung mechanism, by the quarks in the layer of thickness $\lambda $%
. As a first step it is necessary to obtain a quantitative analytical
description of the properties of the electrons outside the quark star.

To find the electron distribution near the quark star surface we follow the
approach of  \citet{Al86}, but also taking into
account the finite temperature effects as discussed by \citet{Ke95}. The
chemical equilibrium implies that the electron chemical potential $\mu
_{\infty }=-V+\mu _{e}$ is constant, where $V$ is the electrostatic
potential per unit charge and $\mu _{e}$ is the electron's chemical
potential. Since far outside the star both $V$ and $\mu _{e}$ tend to zero,
it follows that $\mu _{\infty }=0$ and $\mu _{e}=V$.

For the number density $n_{e}$ of the electrons at the quark star surface we
use the expression \citep{Ke95} 
\begin{equation}
n_{e}\left( z,T\right) =\frac{1}{3\pi ^{2}}\mu _{e}^{3}+\frac{1}{3}\mu
_{e}T^{2}=\frac{1}{3\pi ^{2}}V^{3}+\frac{1}{3}VT^{2},
\end{equation}
where $T$ is the temperature of the electron layer, which can be taken as a
constant, since we assume the electrons are in thermodynamic equilibrium
with the constant temperature quark matter.

The Poisson equation for the electrostatic potential $V\left( z,T\right) $
generated by the finite temperature electron distribution reads \citep{Al86,Ke95}
\begin{equation}
\frac{d^{2}V}{dz^{2}}=\frac{4\alpha }{3\pi }\left[ \left(
V^{3}-V_{q}^{3}\right) +\pi ^{2}\left( V-V_{q}\right) T^{2}\right] ,z\leq 0,
\label{2}
\end{equation}
\begin{equation}
\frac{d^{2}V}{dz^{2}}=\frac{4\alpha }{3\pi }\left( V^{3}+\pi
^{2}VT^{2}\right) ,z\geq 0.  \label{3}
\end{equation}

In Eqs.~(\ref{2})-(\ref{3}), $z$ is the space coordinate measuring height
above the quark surface, $\alpha $ is the fine structure constant and $%
V_{q}/3\pi ^{2}$ is the quark charge density inside the quark matter. The
boundary conditions for Eqs.~(\ref{2})-(\ref{3}) are $V\rightarrow V_{q}$ as 
$z\rightarrow -\infty $ and $V\rightarrow 0$ for $z\rightarrow \infty $. In
the case of the zero temperature electron distribution at the boundary $z=0$
we have the condition $V(0)=(3/4)V_{q}$ \citep{Al86}.

In the following we will be interested mainly in the properties of the
electron gas extending outside the quark surface of the star, described by
Eq.~(\ref{3}). In order to find the exact solution of Eq.~(\ref{3}), we
introduce first a new variable $y=dV/dz$. Then Eq.~(\ref{3}) can be written
in the form 
\begin{equation}
\frac{dy^{2}}{dV}=\frac{8\alpha }{3\pi }\left( V^{3}+\pi ^{2}VT^{2}\right) ,
\end{equation}
immediately leading to the first integral 
\begin{equation}
\frac{dV}{dz}=\sqrt{\frac{4\alpha }{3\pi }}\sqrt{\frac{V^{4}}{2}+\pi
^{2}V^{2}T^{2}},  \label{eqV}
\end{equation}
where an arbitrary integration constant has been set to zero.

The general solution of Eq.~(\ref{eqV}) is given by 
\begin{equation}
V\left( z,T\right) =\frac{2\sqrt{2}\pi T\exp \left[ 2\sqrt{\frac{\alpha \pi 
}{3}}T\left( z+z_{0}\right) \right] }{\exp \left[ 4\sqrt{\frac{\alpha \pi }{3%
}}T\left( z+z_{0}\right) \right] -1},
\end{equation}
where $z_{0}$ is a constant of integration. Its value can be obtained from
the condition of the continuity of the potential across the star's surface,
requiring $V_{I}(0,T)=V\left( 0,T\right) $, where $V_{I}\left( z,T\right) $
is the value of the electrostatic potential in the region $z\leq 0$,
described by Eq.~(\ref{2}). Therefore 
\begin{equation}
z_{0}=\frac{1}{2}\sqrt{\frac{3}{\alpha \pi }}\frac{1}{T}\ln \left[ \frac{%
\sqrt{2}\pi T}{V_{I}\left( 0,T\right) }\left( 1+\sqrt{1+\frac{%
V_{I}^{2}\left( 0,T\right) }{2\pi ^{2}T^{2}}}\right) \right] .
\end{equation}

Hence the electron number density $n_{e}\left( z,T\right) $ is given, at
finite temperatures, by 
\begin{equation}
n_{e}\left( z,T\right) =\frac{2\sqrt{2}\pi }{3}\frac{\exp \left[ 2\sqrt{%
\frac{\alpha \pi }{3}}T\left( z+z_{0}\right) \right] }{\exp \left[ 4\sqrt{%
\frac{\alpha \pi }{3}}T\left( z+z_{0}\right) \right] -1}\left\{ 8\left( 
\frac{\exp \left[ 2\sqrt{\frac{\alpha \pi }{3}}T\left( z+z_{0}\right) \right]
}{\exp \left[ 4\sqrt{\frac{\alpha \pi }{3}}T\left( z+z_{0}\right) \right] -1}%
\right) ^{2}+1\right\} T^{3},T\neq 0.
\end{equation}

In the limit of zero temperature, $T\rightarrow 0$, from Eq.~(\ref{eqV}) we
obtain 
\begin{equation}
V(z)=\frac{a_0}{z+b},  \label{V}
\end{equation}
where $a_0=\sqrt{3\pi /2\alpha }$ and $b$ is an integration constant. $b$ can be determined
from the boundary condition $V(0)=(3/4)V_{q}$, which gives $b=\left( 4a_0/3V_{q}\right) $. Therefore,
in this case for the electron particle number distribution, extending
several thousands fermis above the quark matter surface, we find the
expression: 
\begin{equation}
n_{e}(z)=\frac{1}{3\pi ^{2}}\frac{a^{3}_{0}}{\left( z+b\right) ^{3}},T=0.
\label{ne}
\end{equation}

The variation of the electron number density $n_e(z,T)$ is represented, for
different values of the temperature, in Fig.~7.

In the absence of a crust of the quark star, the electron layer can extend
to several thousands fermis outside the star's surface.

The intensity of the photon beam coming from the thin outer layer of the quark
star decreases due to the energy absorption in the electron layer. The
variation of the intensity of the beam can be obtained by solving the
equation of the radiative transfer \citep{Sha}. For a steady state, the
general solution of the transport equation is by given 
\begin{equation}
I_{\omega }=I_{\omega }\left( 0\right) e^{-\tau _{\omega }}+\int_{0}^{\tau
_{\omega }}j_{\omega }\left( z,T\right) e^{-\left( \tau _{\omega }-\tau
_{\omega }^{\prime }\right) }d\tau _{\omega }^{\prime },  \label{inu}
\end{equation}
where the optical depth $\tau _{\omega }$ is defined as $d\tau _{\omega
}=k_{\omega }\rho dz$, with $k_{\omega }$ the Rosseland mean opacity (giving
the energy absorbed per unit time per unit volume from the beam with
intensity $I_{\omega }$), $I_{\omega }(0)$ is the intensity of the radiation
incident on the electron layer and $j_{\omega }\left( z,T\right) $ is the
emissivity of the electron. Since in the present paper we are interested to study only the effect
of the outer electrons on the quark matter emission, we shall take $%
j_{\omega }\left( z,T\right) \approx 0$, that is we shall neglect the
emissivity of the electron layer. Hence, the variation of the intensity, due
to absorption in the electron layer, of the photon beam emitted by the
quarks at the strange star surface can be described by the following
equation: 
\begin{equation}
I_{\omega }=I_{\omega }\left( 0\right) e^{-\tau _{\omega }}.  \label{inu1}
\end{equation}

In the case of the scattering of photons by free non-relativistic electrons,
the opacity is given by $k_{\omega }=\sigma _{T}n_{e}(z)/\rho $, where $%
\sigma _{T}=(8\pi /3)\left( e^{2}/m_{e}c^{2}\right) ^{2}$ is the Thomson
cross-section \citep{La,Ry79}. This formula is valid only in the case of a non-degenerate gas at
low-temperatures. For systems at high densities and/or temperatures, the
corrections to the Thomson scattering cross-section become important.

For high-temperature non-degenerate electrons the opacity has been
calculated first in \citet{Sa59}. These calculations have been extended to
the semi-degenerate ($T\neq 0$) case by \citet{Ch65} and \citet{BuYu76} (see also \citet{Bo87} and \citet{Ro95}). For a degenerate
electron gas at $T\neq 0$, the Rosseland mean opacity is given by \citep
{BuYu76} 
\begin{equation}
k_{\omega }=\frac{n_{e}\sigma _{T}}{\rho }G^{\deg }\left( T,\eta \right) .
\end{equation}

$G^{\deg }\left( T,\eta \right) $ (the inverse of the Rosseland mean) is a
function of the temperature and of the degeneracy parameter $\eta =\left(
E_{F}-m_{e}c^{2}\right) /kT$, where $E_{F}$ is the Fermi energy. $G^{\deg
}\left( T,\eta \right) $ can be represented in the form \citep{BuYu76} 
\begin{eqnarray}
G^{\deg }\left( T,\eta \right) &=&1.129+0.2965\xi -0.005594\xi ^{2}+\left(
11.47+0.3570\xi +0.1078\xi ^{2}\right) T\nonumber \\
&&+\left( -3.249+0.1678\xi -0.04706\xi
^{2}\right) T^{2},  \label{G}
\end{eqnarray}
where $\xi =\exp \left( 0.8168\eta -0.05522\eta ^{2}\right) $.

The variation of the function $G^{\deg }$ as a function of the parameters $%
\eta $ and $T$ is represented in Fig.~8.

For a non-degenerate electron gas, obeying the Maxwell-Boltzmann
distribution, the Rosseland mean opacity is given by \citep{Sa59} 
\begin{equation}
G^{\textrm{ndeg}}\left( u,T\right) =1+2T+5T^{2}+\frac{15}{4}T^{3}-\frac{1}{5}%
\left( 16+103T+408T^{2}\right) (uT)+\left( \frac{21}{2}+\frac{609}{5}%
T\right) \left( uT\right) ^{2}-\frac{2203}{70}\left( uT\right) ^{3},
\label{ndeg}
\end{equation}
where $u=h\omega /kT$ and the temperature is expressed in units of $%
m_{e}c^{2}$.

In order to obtain a simple estimate of the absorption of the surface
radiation of quarks due to the presence of the electron layer, we consider
first the case of the completely degenerate electron gas at a small
temperature $T$. Complete degeneracy corresponds to $\eta \rightarrow
-\infty $ and $\xi \rightarrow 0$. We also neglect the temperature dependent
terms in Eq.~(\ref{G}), which are not important for small temperatures.
Hence from Eq.~(\ref{G}) we obtain, as a first approximation, $G^{\deg
}\left( 0,-\infty \right) \approx 1.129$.

Therefore the Rosseland opacity of the electron layer at the quark star
surface can be approximated, for low temperatures of the degenerate electron
gas, by the expression 
\begin{equation}
k_{\omega }(z)=\frac{n_{e}\sigma _{T}G^{\deg }\left( 0,-\infty \right) }{%
\rho }=\frac{1}{3\pi ^{2}}\frac{G^{\deg }\left( 0,-\infty \right) \sigma _{T}%
}{\rho }\frac{a^{3}_{0}}{\left( z+b\right) ^{3}}.
\end{equation}

Consequently, for the optical depth of the electron distribution above the
quark star surface we find the expression 
\begin{equation}
\tau _{\omega }=\frac{1}{3\pi ^{2}}G^{\deg }\left( 0,-\infty \right) \sigma
_{T}a^{3}_{0}\int_{0}^{z}\frac{dz}{\left( z+b\right) ^{3}}=\frac{1}{6\pi ^{2}}%
G^{\deg }\left( 0,-\infty \right) \sigma _{T}a^{3}_{0}\left[ \frac{1}{b^{2}}-%
\frac{1}{\left( z+b\right) ^{2}}\right] .  \label{tau0}
\end{equation}

For large $z$, corresponding to the limit $z\rightarrow \infty $, from Eq.~(%
\ref{tau0}) we obtain 
\begin{equation}
\tau _{\omega }=\frac{1}{6\pi ^{2}}G^{\deg }\left( 0,-\infty \right) \sigma
_{T}\frac{a^{3}_{0}}{b^{2}}=\frac{3}{32\pi ^{3/2}}V_{q}^{2}\sqrt{\frac{3}{%
2\alpha }}G^{\deg }\left( 0,-\infty \right) \sigma _{T}.
\end{equation}

At zero temperature the optical depth is proportional to the square of the
value of the quark charge density at the surface of the quark star $V_{q}$.

The absorption of the bremsstrahlung radiation by the electron layer
essentially depends on the quark charge density $V_{q}=\left( 1/3\right)
\left( 2n_{u}-n_{d}-n_{s}\right) $ inside the quark star, where $u$, $d$ and 
$s$ denotes the different quark flavors \citep{Ch98}. The condition of the
charge neutrality requires $V_{q}^{3}=n_{e}^{(int)}$, where $n_{e}^{(int)}$
is the charge density of the electrons inside the star. $n_{e}^{(int)}$ is
related to the baryon number density $n_{b}=\left( n_{u}+n_{d}+n_{s}\right)
/3$ by means of the relation $n_{e}^{(int)}=Y_{e}n_{b}$, where for the
electron abundance $Y_{e}$ we adopt the range $Y_{e}\in \left(
10^{-5},10^{-3}\right) $ \citep{Ch98}. Near the surface of the quark star the
density can be approximated by $\rho \approx 4B\approx 4\times 10^{14}$ g/cm$%
^{3}$. For a strange quark mass of the order of $m_{s}\approx 300$ MeV, for
the baryon number density near the surface of the star we find $n_{b}\approx
7.68\times 10^{6}$ MeV$^{3}$. Hence the corresponding quark charge density
is $V_{q}=197Y_{e}^{1/3}$ MeV, leading to $V_{q}\in (4,20)$ MeV.

The corresponding bremsstrahlung luminosity $L^{\textrm{qs}}$ of the quark
star is given by $L^{\textrm{qs}}=\eta _{eff}4\pi R^{2}L_{Br}=\eta _{eff}4\pi
R^{2}\sigma _{Br}T^{4}$, where $R$ is the radius of the star. For a $T=0$
star and for a degenerate electron layer, 
\begin{equation}
\eta _{eff}=\exp \left( -\tau _{\omega }\right) =\exp \left[ -\left( 3/32\pi
^{3/2}\right) V_{q}^{2}(0)\sqrt{3/2\alpha }G^{\deg }\left( 0,-\infty \right)
\sigma _{T}\right] .
\end{equation}

The variation of the coefficient $\eta _{eff}$ is represented, as a function
of $V_{q}$, in Fig.~9.

For an increasing surface quark charge density $\eta _{eff}$ is a rapidly
decreasing function. Therefore in the case of a large $V_{q}$ the electron
layer significantly decreases the intensity of the bremsstrahlung radiation
emitted by the thin quark layer at the surface of the star.

The spectrum of the $T\neq 0$ quark star essentially depends on the charge
density $V_{I}(0,T)$ at the surface of the star, on the degeneracy
parameter $\eta $ and on the temperature. The numerical values of $%
V_{I}(0,T) $ must be obtained by solving Eq.~(\ref{2}) in the region $z\leq
0 $ for different values of $T$, together with the boundary condition $%
V\rightarrow V_{q}$ for $z\rightarrow -\infty $. Therefore even for the
non-zero temperature quark star, the spectrum of the bremsstrahlung
radiation is determined by the quark charge density inside the star.
However, in order to obtain a simple estimate of the temperature effects on
the optical length of the radiation in the electron layer, we shall assume
that for a suitable range of temperatures the charge density at the surface
can be approximated by $V_{I}(0,T)\approx 14$ MeV \citep{Al86,Ke95}. For this value of the charge density the variation of $%
\eta _{eff}$ as a function of $\eta $ and $T$ is represented in Fig.~10.

Since the increase in the temperature also increases the electron number
density of the electron layer, the absorption of the bremsstrahlung
radiation is much stronger at high temperatures of the semi-degenerate
electron gas.

\section{Discussions and final remarks}

Since the proposal of the existence of strange stars, much effort
has been devoted to find observational properties that could distinguish
strange stars from neutron stars. One of the most common and simplest method
could be to find some specific signatures in the photon emissivity of
strange stars. In the first estimate of the surface photon emissivity of
quark matter \citep{Chmaj} it has been concluded that the most
important photon radiation mechanism is the bremsstrahlung of quarks from a thin layer at the
star's surface, and that the intensity of the radiation is about four orders
of magnitude weaker than the equilibrium black body radiation, although both
have the same temperature dependence.

In the present paper we have reconsidered the electromagnetic emission from the
surface of bare quark stars, by also taking into account the important effect of
the multiple collisions of quarks in the very high density quark-gluon plasma
existing at the surface of the quark stars.  To obtain the bremsstrahlung spectrum of the
photon emission of the quarks we have used several assumptions. The most
important is the adoption of a specific model for the description of
quark-quark elastic scattering, which is essential in obtaining the
bremsstrahlung emissivity. The details of the interactions between quarks
are generally poorly known, and therefore there are a lot of uncertainties
in the description of the strong interactions in a quark-gluon plasma. In
our calculations we have used a classical soft photon approximation, neglecting
quantum effects. We
have generally assumed that the quarks are semi-degenerate,
with temperature $T>0$ and $T<<E_{F}$, but
with the kinetic energies of the interacting particles of the order of the Fermi energy.

In order to give a complete description of the quark-quark bremsstrahlung we
have considered, in a systematic way, the effects of the dense medium on the
radiation spectrum. A dense medium in which many collisions occur will
reduce the bremsstrahlung radiation, and as high the density is, as lower is
the bremsstrahlung emission. On the other hand the emission of photons with
frequency smaller than the critical LPM frequency is completely suppressed. $%
\omega _{LPM}$ depends on the QCD coupling constant and on the density at
the quark star's surface. For strange stars $\omega _{LPM}$ is of the order
of $\omega _{LPM}\approx 5-6$ MeV, so photons with $\omega <\omega _{LPM}$
can not be emitted from the surface of the star. Hence only low frequencies are
suppressed due to the LPM effect. On the other hand the multiple collision effect
reduces with one order of magnitude the bremsstrahlung coefficient $\sigma _{Br}$,
making the soft photon radiation from the quark star $5$ orders of magnitude weaker than the black body
radiation.

A second effect, which essentially influences the radiation of photons by
quark stars, is the absorption of the radiation in the electron layer formed
outside the quark star. The electron layer could extend to several
hundreds or thousands fermis, the electron number density being an increasing
function of the temperature. The high-density semi-degenerate electron layer
at the surface of a strange star
plays an important role in the propagation of the electromagnetic radiation
from the star's surface. The optical depth of the degenerate or semi-degenerate 
electron layer depends on the electron number density, temperature, the
degeneracy parameter and the charge density at the surface of the quark
star. For high surface charge densities the intensity of the quark
bremsstrahlung radiation is significantly reduced by absorption in the
electron layer. For the low-temperature star, with $T<<E_F$, for a surface potential
$V_q\approx 20$ MeV, the electron layer reduces the intensity of the radiation by an
order of magnitude. In this case the combined effects of the multiple collisions
and absorption in the electron layer give $\sigma _{Br}/\sigma _{SB}\sim
10^{-6}$. Hence the energy radiated away by the quark star could be $6$ orders
of magnitude smaller than the equilibrium black body radiation.

For the sake of comparison we also present the expression for the photon
emission rate of an equilibrated, hot QCD plasma, at zero chemical
potential, with the LPM effect fully included. The spontaneous emission rate 
$\left( 2\pi \right) ^{3}dI_{\gamma }/d^{3}k$ of photons with a given
momentum $\vec{k}$ can be represented, for two flavor QCD, as (Arnold et al.
2001, Arnold et al. 2002)
\begin{equation}
\left( 2\pi \right) ^{3}\frac{dI_{\gamma }}{d^{3}k}=\frac{40\pi \alpha
\alpha _{s}}{9}T^{2}\frac{n_{f}\left( \omega \right) }{\omega }\left[ \ln 
\frac{T}{m_{\infty }}+C_{tot}\left( \frac{\omega }{T}\right) \right] ,
\label{ar1}
\end{equation}
where $n_{f}\left( \omega \right) $ is the Fermi-Dirac distribution
function, $n_{f}\left( \omega \right) =\left[ \exp \left( \omega /T\right) +1%
\right] ^{-1}$, $m_{\infty }$ is the thermal quark mass, given by $m_{\infty
}^{2}=4\pi \alpha _{s}T^{2}/3$, and
\begin{equation}
C_{tot}\left( \frac{\omega }{T}\right) \equiv \frac{1}{2}\ln \frac{2\omega }{%
T}+C_{2\leftrightarrow 2}\left( \frac{\omega }{T}\right) +C_{brem}\left( 
\frac{\omega }{T}\right) +C_{annih}\left( \frac{\omega }{T}\right) ,
\label{ar2}
\end{equation}
with $C_{2\leftrightarrow 2}\left( x\right) \approx
0.041x^{-1}-0.3615+1.01\exp (-1.35x)$ and $C_{brem}\left( x\right)
+C_{annih}\left( x\right) =\sqrt{1+N_{f}/6}$ $\left[ 0.548\ln %anisia
(12.28+1/x)/x^{3/2}+0.133x/\sqrt{1+x/16.27}\right] $. For $\omega /T<2.5$,
bremsstrahlung becomes the most important process, while for $\omega /T>10$,
pair annihilation dominates. The LPM effect suppresses both bremsstrahlung and pair
annihilation processes, but the suppression is not severe ($35\%$ or less),
except for photons with $\omega <2T$ or very hard pair annihilation. 

In the limit of small temperatures, $T\rightarrow 0$ and $%
x\rightarrow \infty $. Hence $C_{2\leftrightarrow 2}\left( x\right)
\rightarrow -0.3615$ and $C_{brem}\left( x\right) +C_{annih}\left( x\right)
\rightarrow 0.133\left( \omega /T\right) ^{1/2}$. By neglecting the
logarithmically divergent terms, the integration over all possible ranges of
frequencies in Eq. (\ref{ar1}) gives  
\begin{equation}
I_{\gamma }\sim \chi _{Br}^{(QCD)}T^{4},
\end{equation}
where $\chi _{Br}^{(QCD)}\approx 5\pi \alpha \alpha _{s}\left[ 3\sqrt{%
1+N_{f}/6}\left( 4-\sqrt{2}\right) \pi ^{-3/2}\xi \left( 5/2\right)
/4-0.3615/3\right] /9$, with $\xi \left( s\right) $ is the Riemann zeta
function $\xi \left( s\right) =\sum_{k=1}^{\infty }k^{-s}$. Therefore the
perturbative QCD approach gives, in the low temperature limit, the same
temperature dependence of the bremsstrahlung spectrum as in Eq. (\ref{arbr}%
). However, we must point out that Eqs.(\ref{ar1}) and (\ref{ar2})  have
been derived for a temperature of the quark-gluon plasma higher than $250$
MeV, since the temperature of the fireball formed in heavy ion collisions is
of the order of $450$ MeV (Renk 2003). 

Recently, it has been argued that the strange quark matter in the
color-flavor locked(CFL) phase of QCD, which occurs for large gaps $(\Delta
\sim 100MeV)$, is rigorously electrically neutral, despite the unequal quark
masses, and even in the presence of electron chemical potential \citep{Al98,Al99,Ra01}.

However, \citet{PaUs02} pointed out that for sufficiently large $m_{s}$
the low density regime is rather expected to be in the ''2-color-flavor
Superconductor'' phase in which only $u$ and $d$ quarks of two color are
paired in single condensates, while the ones of the third color and $s$ quarks
of all three colors are unpaired. In this phase, electrons are present. In
other words, electrons may be absent in the core of strange stars but
present, at least, near the surface where the density is lowest.
Nevertheless, the presence of CFL effect can reduce the electron density at
the surface and hence increases the bremsstrahlung emissivity.

In conclusion, in the present paper we have pointed out two effects that
could significantly reduce or even fully suppress the bremsstrahlung radiation from the surface of
quark stars: the significant decrease of the intensity due to the multiple collisions in a dense
medium and the absorption by the outer electron layer of the stars. The possible observational
significance for the detection of quark stars of these effects will be considered in a future publication.

\section*{Acknowledgments}

The authors would like to thank to the anonymous referee, whose comments helped
to improve the manuscript. We are grateful for the useful comments of Prof. V. V. Usov.
This work is supported by a RGC grant of Hong Kong Government.




\clearpage

\begin{figure}
\plotone{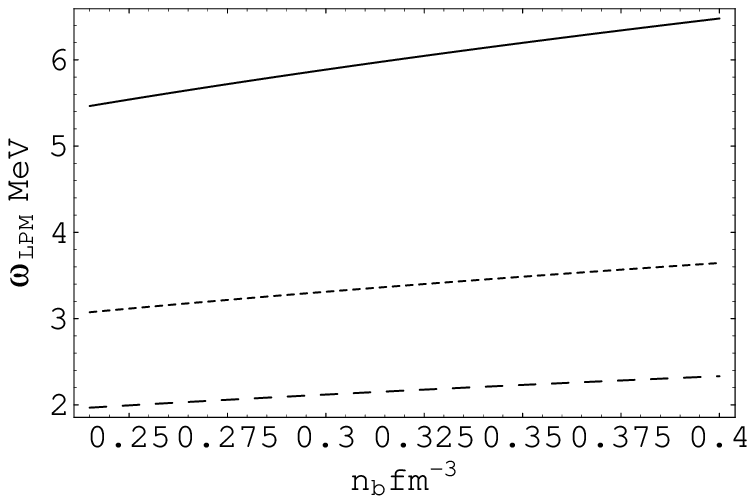}
\caption{Variation, as a function of the quark number density $n_b$, of the LPM
frequency $\omega _{LPM}$ for different values of the strong
coupling constant $\alpha _{s}$: $\alpha _{s}=0.3$ (solid
curve), $\alpha _{s}=0.4$ (dotted curve) and $\alpha %
_{s}=0.5 $ (dashed curve). \label{FIG1}}
\end{figure}

\begin{figure}
\plotone{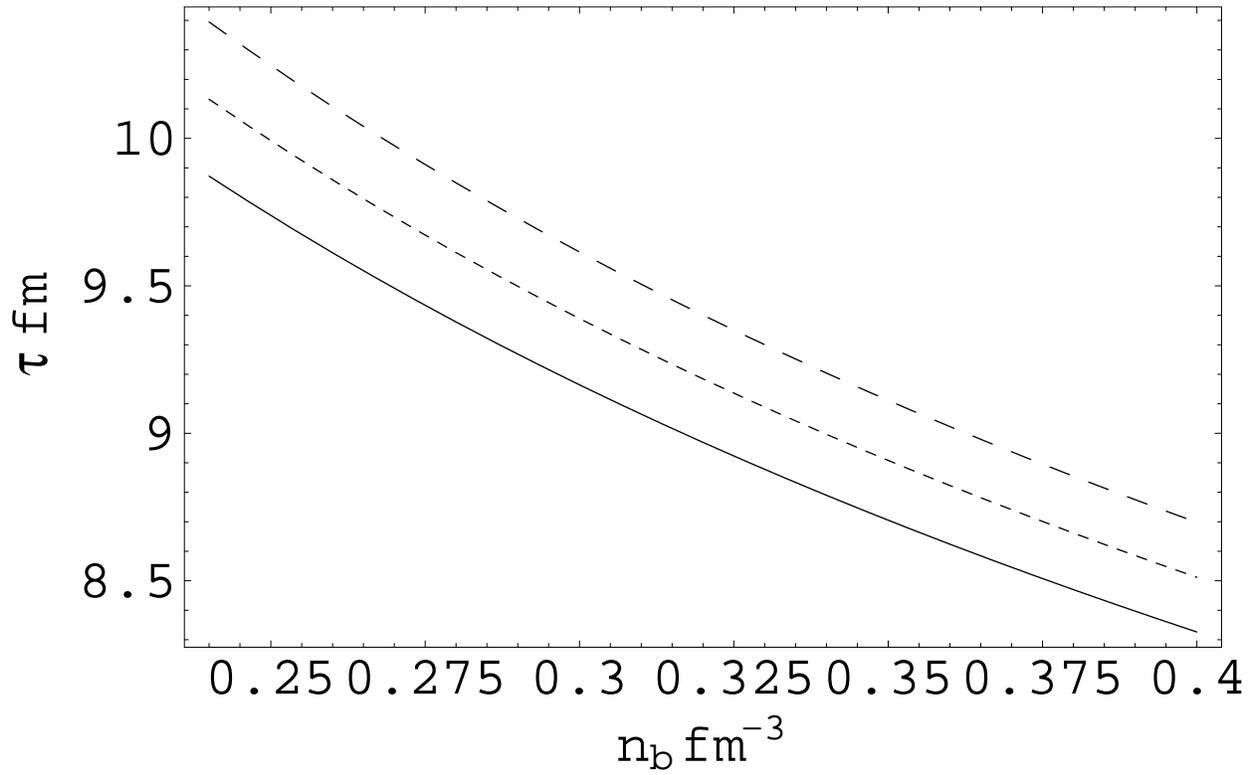}
\caption{Variation, as a function of the quark number density, of the mean
collision time $\tau $ for different values of the temperature of
the surface of the strange star: $T=0.10$ MeV (solid curve), $T=5$ MeV
(dotted curve) and $T=10$ MeV (dashed curve). \label{FIG2}}
\end{figure}

\begin{figure}
\plotone{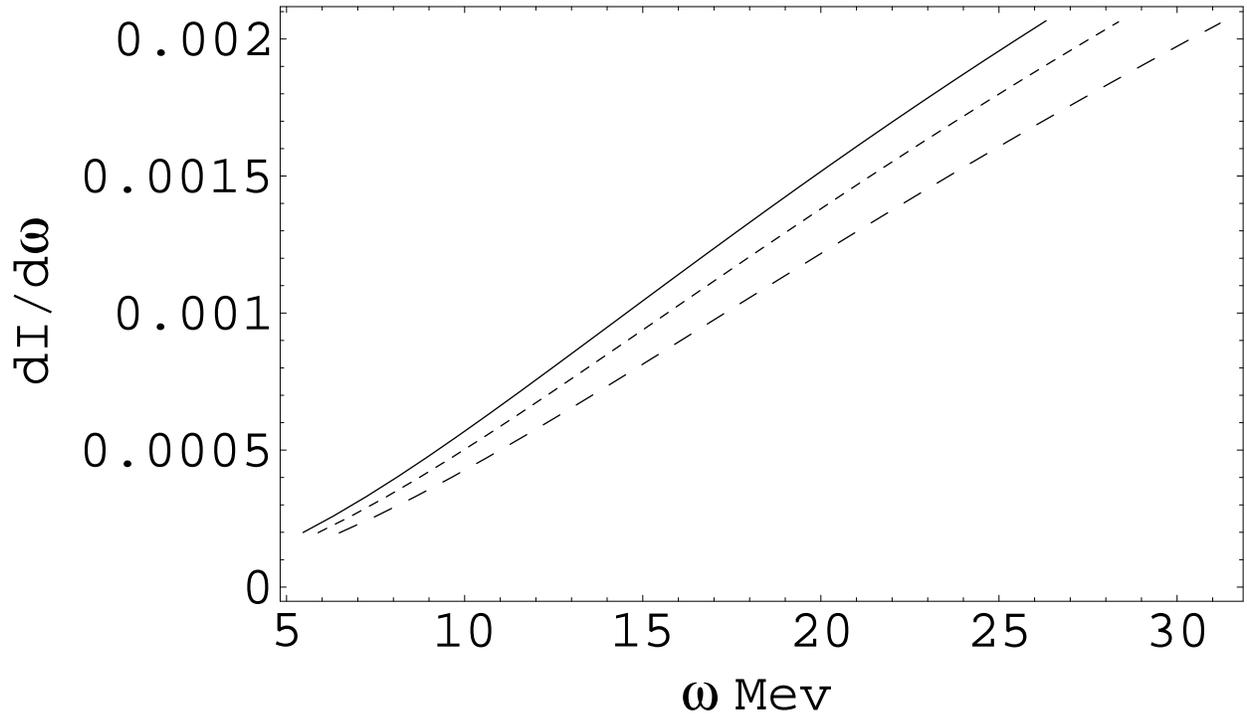}
\caption{Frequency distribution $dI/d\omega $ of the bremsstrahlung
radiation emitted due to quark-quark collisions at the surface of the
strange star, for different values of the surface density: $n_b=0.24 \textrm{fm%
}^{-3}$ (solid curve), $n_b=0.30 \textrm{fm}^{-3}$ (dotted curve) and $n_b=0.4 
\textrm{fm}^{-3}$ (dashed curve). In all cases we have considered a temperature of $T=5$ MeV
of the strange matter. For the QCD coupling constant we have chosen the value $%
\alpha _s=0.4$. \label{FIG3}}
\end{figure}

\begin{figure}
\plotone{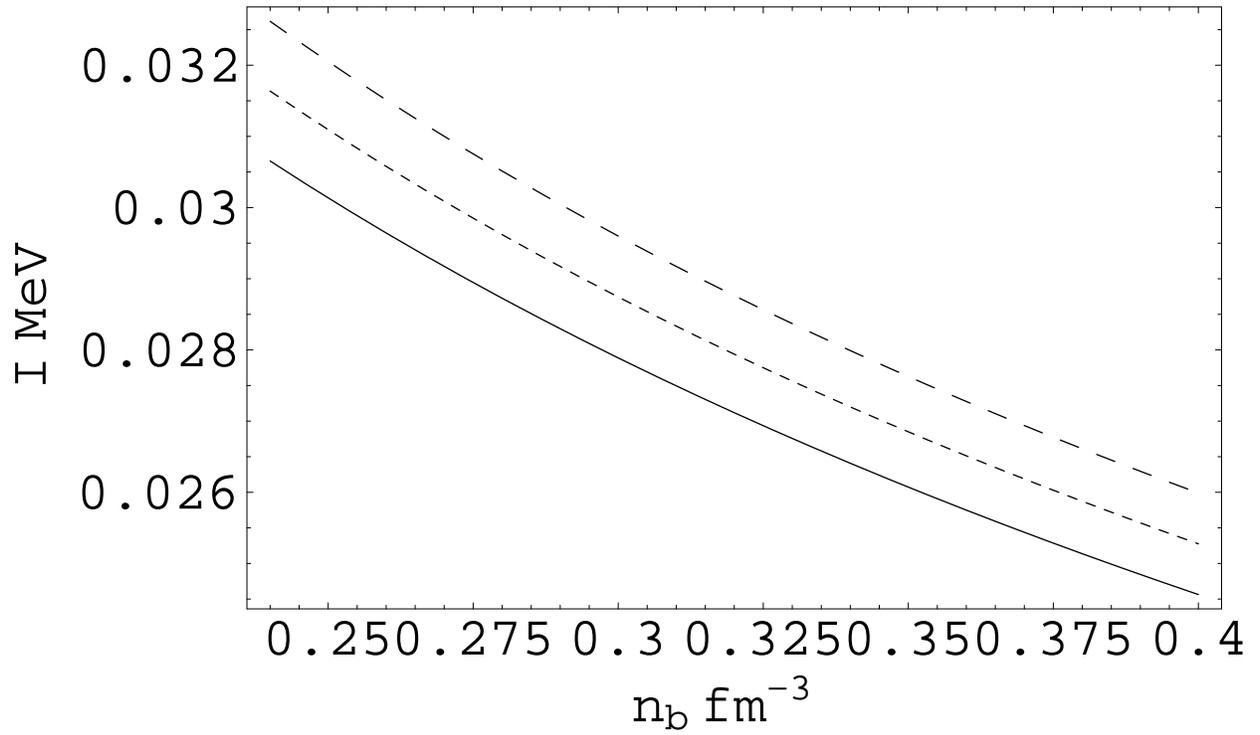}
\caption{Variation of the total intensity $I$ of the bremsstrahlung radiation emitted by a
single quark as a function of the collision time $\tau $. We assumed that the
velocity of the quarks is the Fermi velocity and the QCD coupling
constant is $\alpha _s=0.4$. \label{FIG4}}
\end{figure}

\begin{figure}
\plotone{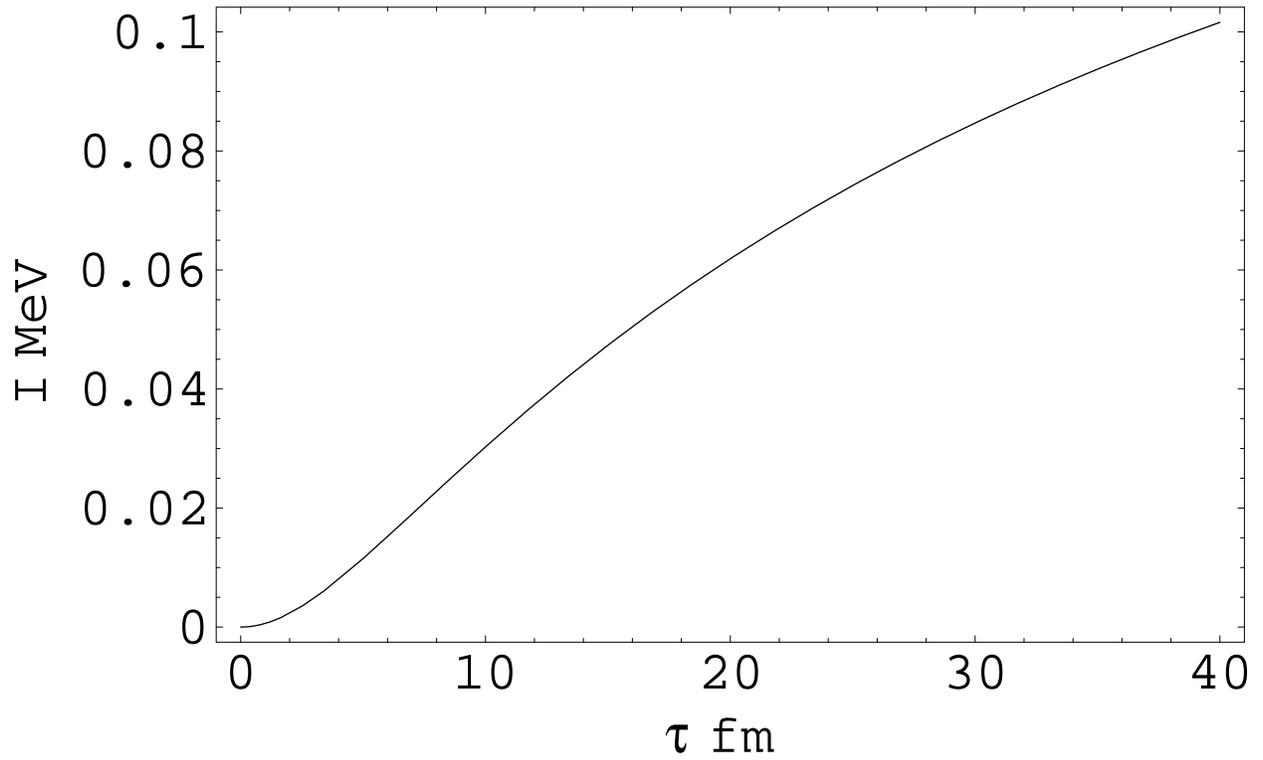}
\caption{Total intensity $I$ of the bremsstrahlung radiation emitted by a
single quark as a function of the strange star's surface density $n_b\textrm{
fm}^{-3}$, for different values of the temperature: $T=0.10$ MeV (solid
curve), $T=5$ MeV (dotted curve) and $T=10$ MeV (dashed curve). In all cases
we have taken $\alpha _s=0.4$. \label{FIG5}}
\end{figure}

\begin{figure}
\plotone{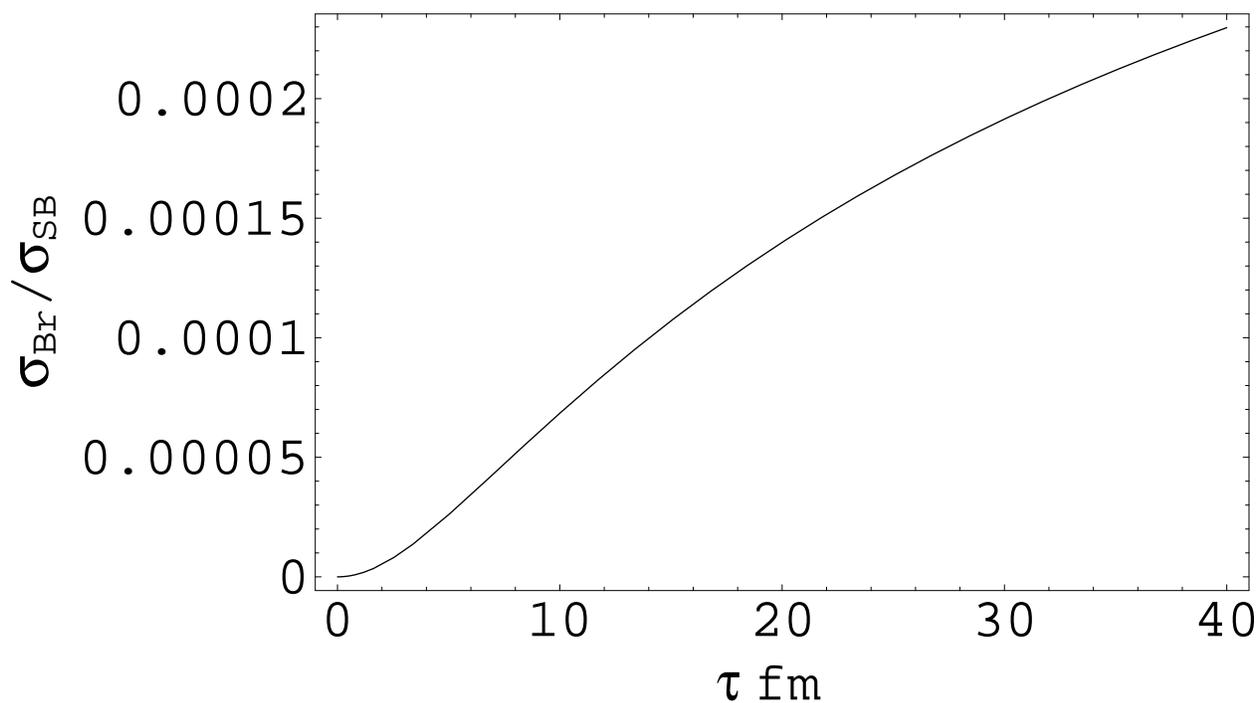}
\caption{Variation of the ratio $\sigma _{Br}/\sigma _{SB}$ of the
bremsstrahlung emissivity of the quark matter and of the Stefan-Boltzmann constant,
as a function of the mean collision time $\tau $ at the quark star surface.
The thickness of the emitting quark layer is $\lambda =5$ fm. \label{FIG6}}
\end{figure}

\begin{figure}
\plotone{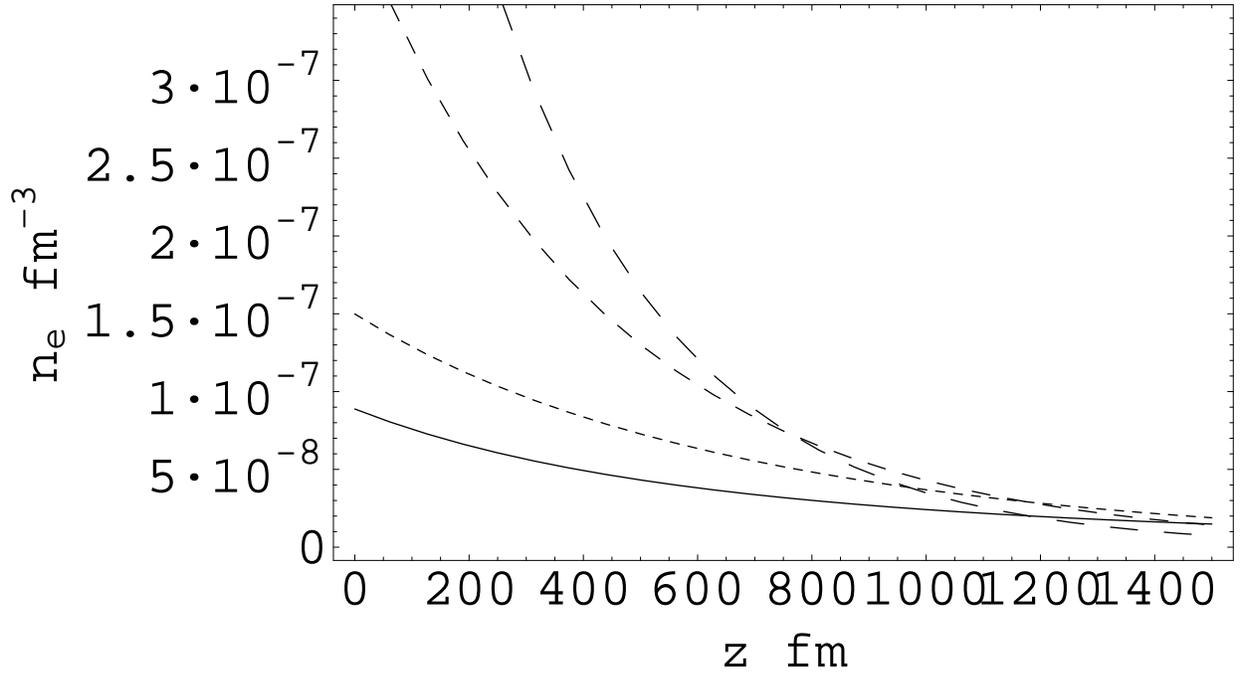}
\caption{Variation of the electron number density $n_e$ as a function of
the distance $z$ (fm) for different values of the temperature: $T=0$ (solid
curve), $T=1.5$ MeV (dotted curve), $T=2.5$ MeV (dashed curve) and $T=3.5$
MeV (long dashed curve). For the $T=0$ case we have chosen $V_q=20$ MeV,
while for $T\neq 0$, $V_{I}(0,T)=14$ MeV. \label{FIG7}}
\end{figure}

\begin{figure}
\plotone{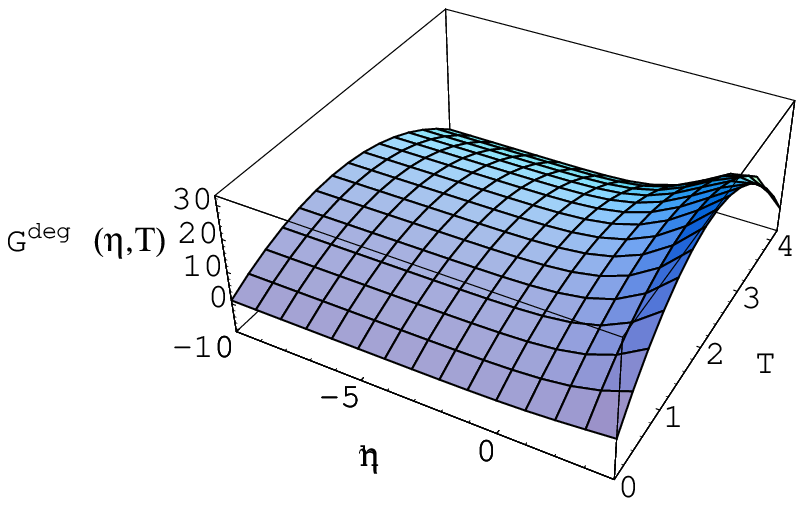}
\caption{Variation of the function $G^{\deg }\left( T,\eta \right) $
for $\eta \in \left( -10,+4\right) $ and $T\in \left(0,
4.5m_{e}c^{2}\right) $. \label{FIG8}}
\end{figure}

\begin{figure}
\plotone{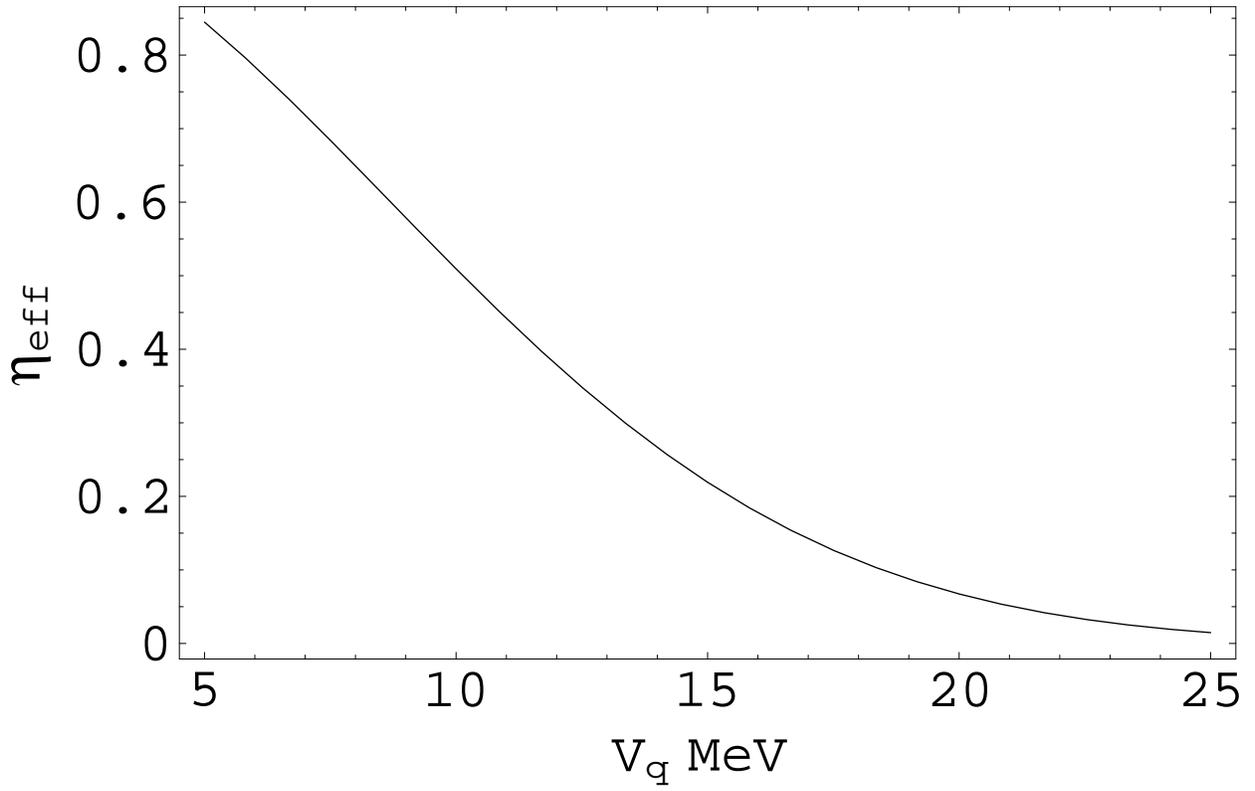}
\caption{Variation of $\eta _{eff}$ as a function of the strange
quark surface potential $V_{q}$. \label{FIG9}}
\end{figure}

\begin{figure}
\plotone{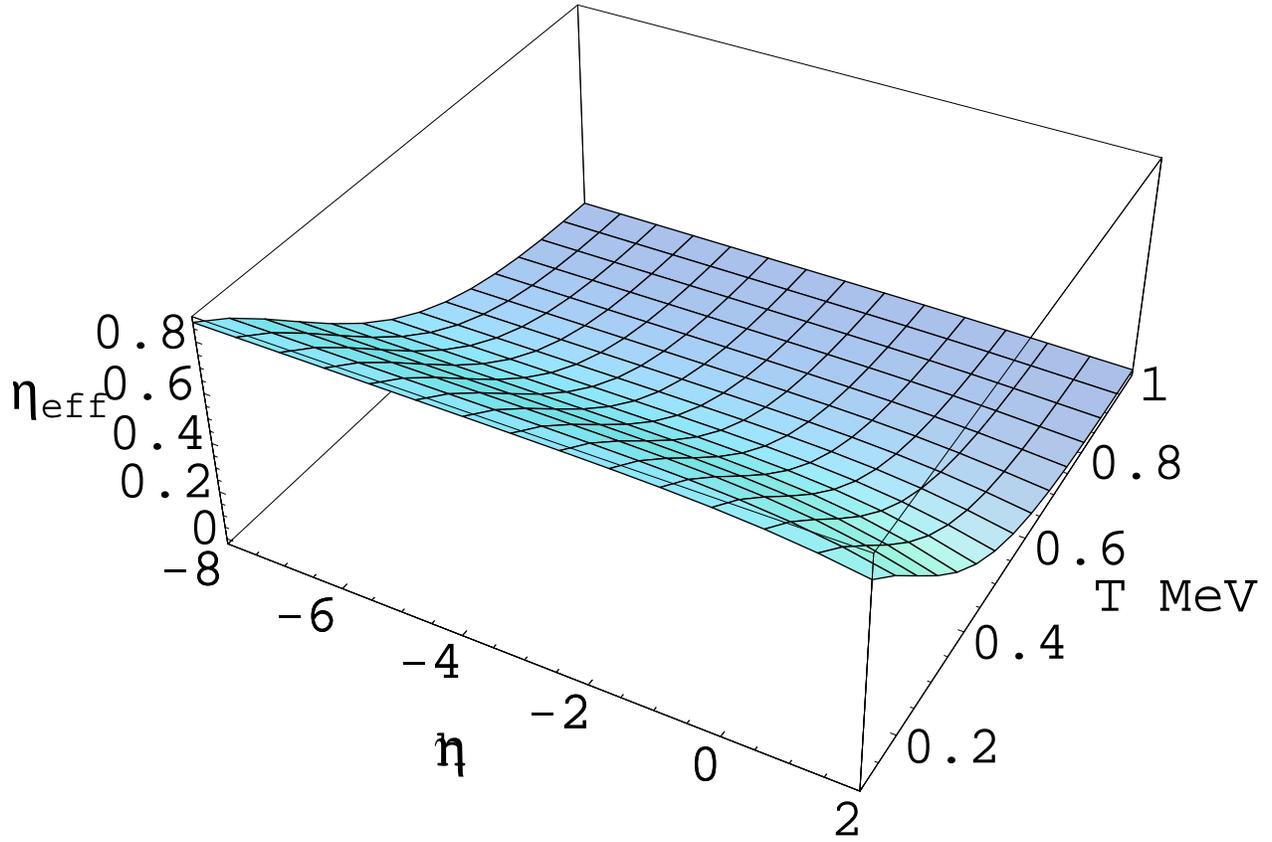}
\caption{Variation of $\eta _{eff}$ as a function of the degeneracy
parameter $\eta $ and of the temperature $T$ for the case of a
partially degenerate electron gas, for a quark surface charge density
potential $V_{I}=14$ MeV. \label{FIG10}}
\end{figure}
\end{document}